\documentclass[5p,twocolumn,times]{elsarticle}
\usepackage{ecrc}

\volume{00}

\firstpage{1}

\journalname{Journal of Parallel and Distributed Computing}

\runauth{C. Rivera et al.}

\jnltitlelogo{JPDC}

\usepackage{etoolbox}
\patchcmd{\pprintMaketitle}
 {\ifvoid\absbox\else\unvbox\absbox\par\vskip10pt\fi}
 {\ifvoid\absbox\else\clearpage\unvbox\absbox\par\vskip30pt\fi}
 {}{}
\patchcmd{\pprintMaketitle}
 {\hrule\vskip12pt}
 {}
 {}{}
\patchcmd{\pprintMaketitle}
 {\hrule\vskip12pt}
 {}
 {}{}
\appto{\pprintMaketitle}{\clearpage}

\usepackage{lineno}
\modulolinenumbers[5]
\usepackage{hyperref}
\usepackage{amsmath,amssymb,amsfonts}
\usepackage{algcompatible} % for disabling number in algorithm
\usepackage{graphicx}
\usepackage{textcomp}
\usepackage{xcolor}
\usepackage{soul}
\usepackage{color}
\usepackage{graphicx}
\usepackage{caption}
\usepackage{subcaption}
\usepackage{comment}
\usepackage{algorithm2e}
\usepackage[normalem]{ulem}
\usepackage{booktabs}
\usepackage{todonotes}
\usepackage{picins}
\usepackage[final]{pdfpages}

\PassOptionsToPackage{protrusion, expansion}{microtype}
\PassOptionsToPackage{hyphens}{url}\usepackage{hyperref}
\newcommand\revision[1]{\textcolor{black}{#1}}
\newcommand{\comm}[1]{}
\newcommand\black[1]{\textcolor{black}{#1}}

\begin{document}

%\pagestyle{empty}
%\pagenumbering{plain}
%\pagenumbering{gobble}

\begin{frontmatter}

\title{\textsc{TSM2X}: High-Performance Tall-and-Skinny Matrix-Matrix Multiplication on GPUs}

\author[1]{Cody Rivera\fnref{fn1}}
\author[2]{Jieyang Chen\fnref{fn1}}
\author[3]{Nan Xiong}
\author[4]{Jing Zhang}
\author[5]{Shuaiwen Leon Song}
\author[6,1]{Dingwen Tao\corref{cor1}}
\ead{dingwen.tao@wsu.edu}

\fntext[fn1]{Cody Rivera and Jieyang Chen have contributed equally to this work.}
\cortext[cor1]{Corresponding author}

\address[1]{The University of Alabama, Tuscaloosa, AL 35487, USA}
\address[2]{Oak Ridge National Laboratory, Oak Ridge, TN 37830, USA}
\address[3]{The University of Tennessee, Knoxville, TN 37996, USA}
\address[4]{University of Colorado Colorado Springs, CO 80918, USA}
\address[5]{The University of Sydney, NSW 2006, Australia}
\address[6]{Washington State University, Pullman, WA 99164, USA}

\pagenumbering{gobble}
\begin{abstract}
Linear algebra operations have been widely used in big data analytics and scientific computations. Many works have been done on optimizing linear algebra operations on GPUs with regular-shaped input. However, few works focus on fully utilizing GPU resources when the input is not regular-shaped. Current optimizations do not consider fully utilizing the memory bandwidth and computing power; therefore, they can only achieve sub-optimal performance. 
In this paper, we propose two efficient algorithms---\texttt{TSM2R} and \texttt{TSM2L}---for two classes of tall-and-skinny matrix-matrix multiplications on GPUs. Both of them focus on optimizing linear algebra operation with at least one of the input matrices is tall-and-skinny. 
Specifically, \texttt{TSM2R} is designed for a large regular-shaped matrix multiplying a tall-and-skinny matrix, while \texttt{TSM2L} is designed for a tall-and-skinny matrix multiplying a small regular-shaped matrix.
We implement our proposed algorithms and test on several modern NVIDIA GPU micro-architectures. Experiments show that, compared to  the current state-of-the-art works, (1) \texttt{TSM2R} speeds up the computation by \black{1.6x on average} and improves the memory bandwidth utilization and computing power utilization by \black{18.1\%} and \black{20.5\% on average}, respectively, when the regular-shaped matrix size is relatively large or medium; and (2) \texttt{TSM2L} speeds up the computation by \black{1.9x on average} and improve\black{s} the memory bandwidth utilization by up to\comm{55\%} \black{9.3\% on average} when the regular-shaped matrix size is relatively small.
\end{abstract}


\end{frontmatter}

\pagenumbering{arabic}

\section{Introduction}

Matrix-matrix multiplication (GEMM) has been one of the most extensively used linear algebra operations in big data analytics and scientific computations. Due to many factors (such as algorithms, input data, etc.) the size or shape of input matrices for GEMM usually varies when it is used in different applications.
For example, many modern highly scalable scientific simulation packages in the field of fluid dynamics, such as Finite Element Method (FEM) \black{simulations,} need to compute many GEMMs with small-sized input \black{matrices}. Artificial neural networks (ANN) involve using GEMM with small to medium input matrices. Matrix decompositions uses GEMM with large-sized input matrices \cite{magma,chen2016greenla,tan2014survey,tan2015investigating}. Thus, besides large-sized input, which has already been extensively optimized during the past decades, GEMM with small to medium sized input has also drawn much attention to recent researchers. For instance, Dong et al. \cite{dong2016magma} proposed MAGMA-Batched, which aims to batch small input matrices into larger ones in order to utilize the highly optimized implementations for large input size on GPUs. 
Heinecke et al. \cite{heinecke2016libxsmm} proposed to speed up GEMM with small input using architecture and instruction level optimization on modern CPU architectures. 

Although previous works have focused on optimizing GEMM with different matrix sizes, most of them only assume that the input matrices are \textit{regular-shaped}. 
In other words, the size mentioned in their works usually refers to both dimensions of the input matrix. 
For example, a small matrix means both of its width and height are small and their magnitudes are close to each other. 
When the dimensions of the input matrices have significant difference, we consider them as \textit{irregular-shaped} inputs.
In particular, many irregular-shaped inputs involve \textit{tall-and-skinny} matrices, in which their widths are significantly smaller then their heights.
\black{Although few works have been done to study and optimize GEMM with tall-and-skinny input}, \black{this input case} has been widely used in many applications \cite{chen2019tsm2}.
For instance, recent highly optimized K-means implementations \cite{dhillon2004kernel,k-means} use GEMM as their core computation, and the input size is mostly tall-and-skinny. This is because the number of centroids is usually far less than the number of input data points.
Moreover, when GEMM is used for encoding checksums for many algorithm-based fault tolerance applications \cite{chen2016online, huang1984algorithm, chen2016gpu, chen2018fault, chen2016tao, chen2016wu, wu2017silent, liang2017correcting,wu2011fault,chen2019fault, tao2018improving}, the input involves a tall-and-skinny checksum weight matrix.

%Moreover, for deep neural networks with a large number of parameters (e.g., ResNet \cite{he2016deep}), 
%Moreover, deep neural networks usually need to train a large number of parameters (e.g., ResNet \cite{he2016deep}).
%limited by the GPU/CPU memory space the batch size for each training epoch is usually small, so the GEMM involved its computation becomes tall-and-skinny, in which input matrix has one significant small dimension. 
%The limited GPU/CPU memory space constraints the batch size for each training epoch. In this case, the matrix-matrix multiplication becomes tall-and skinny and the input matrix has one significant small dimension.
%Also, for the core part of widely used k-means clustering algorithm also uses tall-and-skinny GEMM for calculating distances between each data point to each centroid \cite{k-means}. 

%However, optimization techniques used for optimizing regular size GEMM are not usually suitable for optimizing GEMM with none regular sized input, so they do not deliver the best possible performance, which fully utilizes underlying hardware resources. 
Previous efforts made for optimizing GEMM with regular-shaped input may not work for non-regular shaped input. For instance, Chen et al. \cite{chen2018fault} illustrates that calculating GEMM with tall-and-skinny input using \black{the} vendor’s highly optimized linear algebra library (e.g., cuBLAS \cite{cublas}) is slower than disassembling the tall-and-skinny input matrix into several vectors and then applying matrix-vector multiplications.
However, it can be easily seen that even with this workaround the computation is not efficient, since elements in input matrices are accessed by the GPU more times than necessary. Although the performance can be optimized by grouping many tall-and-skinny input matrices into large ones similar to the approach proposed, there are cases where this grouping approach is not feasible. For example, tall-and-skinny input matrices may be generated one at a time from a producer process in user’s workflow. Grouping several of them into a large matrix requires extended waiting time, which is not applicable for time-sensitive applications. On the other hand, the memory space may limit the total number of matrices that can fit into the memory at the same time, if the input matrices are large (e.g., multiplication of regular-shaped large and tall-and-skinny matrices).

In this work, we target on optimizing the computation of GEMM
with tall-and-skinny input on the GPU platform since many applications that use GEMM are deployed on GPUs. So, our optimization greatly benefits those applications. The key insight of our work is that the computation characteristic of GEMM on modern computing systems is not always unchanged as we change the shape of input matrices.
For example, when the sizes of regular-shaped matrices are large (i.e., $m \simeq k \simeq n \gg 1$ for an $m \times k$ matrix multiplying an $k \times n$ matrix), the compute-to-load ratios of each element in the input matrices are \black{$O(m) \approx O(n)$}.
So, the \textit{regular-shaped} GEMM operations are usually compute-bound especially for large matrices.
However, when the input is tall-and-skinny (i.e., $m \simeq k \gg n$ or $m \gg k \simeq n$), the average compute-to-load ratio is reduced \black{to} around $O(1)$.
Moreover, when $k$ is very small (i.e., $m \gg k \simeq n$), each GPU thread would not perform enough workload to hide latency and hence low occupancy.
Therefore, depending on the relationship between $m$, $k$, and $n$, and the performance characteristics of GPUs, the computation can be compute-bound, memory-bound, or latency-bound. 
Specifically, when
(1) $m \simeq k \gg n$, as $n$ gets larger, it moves toward compute-bound; 
(2) $m \simeq k \gg n$, as $n$ gets smaller, it moves toward memory-bound; and
(3) $m \gg k \simeq n$, it moves toward latency-bound.
To optimize GEMM with tall-and-skinny input, it is critical to design a computation algorithm that considers all compute-bound, memory-bound, and latency-bound cases.

The main contributions of this paper include:
\begin{itemize}
\item We study the limitation of current state-of-the-art GEMM implementations with tall-and-skinny inputs (i.e., $m \simeq k \gg n$ or $m \gg k \simeq n$). 
With benchmarking, we find that the under-utilization of GPU resources is the main reason for performance degradation when the input is tall-and-skinny.
\item To handle a broad spectrum of tall-and-skinny inputs for GEMM on GPUs, we design two classes of algorithms with optimizations focusing on different tall-and-skinny input cases: 
(1) \texttt{TSM2R} is designed to handle a large regular-shaped matrix multiplying a tall-and-skinny matrix (i.e.,  $m \simeq k \gg n$); 
(2) \texttt{TSM2L} is designed to handle a tall-and-skinny matrix multiplying a small regular-shaped matrix (i.e., $m \gg k \simeq n$).
\item We present a performance model for \texttt{TSM2R} and compare it with our evaluation performance results. Moreover, we examine the inadequacies of the model for \texttt{TSM2L} and further improve it based on our observations.
\item We carefully implement \texttt{TSM2R} and \texttt{TSM2L} using CUDA C\footnote{The TSM2X code is available at \url{https://github.com/codyjrivera/tsm2x-imp}.} and evaluate them on four generations of NVIDIA GPUs including Kepler, Maxwell, Pascal, and Volta. Experiments show that our \texttt{TSM2R} and \texttt{TSM2L} can achieve \black{1.6x} and \black{1.9x} speedups, respectively, \black{on average} with different tall-and-skinny inputs, compared to the state-of-the-art GEMM library cuBLAS.
\end{itemize}

The rest of this paper is organized as follows. In Section \ref{backgrouds}, we give a formal definition of tall-and-skinny matrix and show some preliminary benchmark results of the GEMM with tall-and-skinny matrix using cuBLAS.
In Section \ref{design}, we propose our detailed design of \texttt{TSM2R} and \texttt{TSM2L} for two different kinds of tall-and-skinny inputs. In Section \ref{experiments}, we present our evaluation results. \black{In Section \ref{related}, we examine related works for tall-and-skinny inputs.} In Section \ref{conclusion}, we conclude the paper.
\section{Background}
\label{backgrouds}

\subsection{Tall-and-Skinny Input for GEMM}
\label{problem}
In this work we restrict our scope to handle irregular-shaped inputs that involve tall-and-skinny matrices.
The tall-and-skinny input size means that, for the two input matrices, at least one matrix is tall-and-skinny (i.e., one dimension is significantly smaller than the other). 
For example, either (i) input matrix A with size $20480 \times 20480$ and matrix B with size $20480 \times 2$ or (ii) input matrix A with size $20480 \times 2$ and matrix B with size $2 \times 2$ is considered as tall-and-skinny input in our work.
Tall-and-skinny matrices are a typical class of matrices that can be found in irregular-shaped inputs for GEMM.
In this paper, we focus on optimizing GEMM with (i) one large regular input matrix and one tall-and-skinny input matrix and (ii) one tall-and-skinny input matrix and one small regular input matrix.
In this paper, for the first case, we let matrix A be the larger input matrix ($m \times k$) and matrix B ($k \times n$) be the tall-and-skinny input matrix, where $m \simeq k \gg n$; for the second case, we let matrix A be the tall-and-skinny input matrix ($m \times k$) and matrix B ($k \times n$) be the smaller input matrix, where $m \gg k \simeq n$.
We choose these input sizes and shapes because we believe they can expose most of the challenges in processing all kinds of tall-and-skinny input, so the design idea and optimization techniques introduced in this paper can be easily applied to other cases with slight modification. Also, for simplicity's sake, we choose to let the larger matrix in (i) and smaller matrix in (ii) to be square-shaped in most of our experiments. Our optimization can work with non-square input as well with similar effects.

\subsection{cuBLAS}
One of the most commonly used standard linear algebra libraries optimized for \revision{the} GPU is the cuBLAS library developed by NVIDIA. cuBLAS is the core computing library of many big data and scientific computing applications. 
For example, it is the GPU computing library for MAGMA heterogeneous linear algebra library \cite{tdb10, tnld10, dghklty14}, cuLA library \cite{cula}, and cuDNN deep learning library \cite{cudnn}. 
With NVIDIA's deep optimization, the cuBLAS library is able to provide state-of-the-art performance in many use cases. 
For example, with large regular-shaped input matrix, their GEMM implementation is able to achieve near peak GPU performance  \cite{cublas-banchmark}. 

However, we found that the GEMM subroutine is not fully optimized with certain input matrix sizes \cite{chen2016online}. 
For example, with inputs that involve tall-and-skinny matrices, the GEMM operation in current best implementation (cuBLAS 9.0 running on NVIDIA Tesla K40c GPU) uses less than 10\% of the theoretical peak memory bandwidth on average with $n = 2$ (as demonstrated in Figure \ref{k40-mem} (a)-(b)). 
When $n=16$, the same GEMM operation uses less than 20\% of the theoretical peak memory bandwidth on average (as demonstrated in  Figure \ref{k40-mem} (g)-(h)). 
The resource utilization is even lower with larger input dimensions. 
By comparing the two input sizes, it can be seen for input with smaller $n$ values, the computation utilizes higher memory bandwidth (close to memory bound). 
On the other hand, for input with larger $n$ values, the computation utilizes higher computing power (close to compute bound). 
However, since we are unable to analyze the GEMM implementation in the \revision{closed-source} cuBLAS library, it is hard to tell \revision{its exact computational characteristics.}

\section{Design Methodologies}
\label{design}
To handle the GEMM with two different classes of tall-and-skinny inputs on GPUs described in Section \ref{problem}, we design two efficient algorithms: \texttt{TSM2R} and \texttt{TSM2L}.
\texttt{TSM2R} is \black{designed} to handle inputs with one large-to-medium regular-shaped matrix and one tall-and-skinny matrix, while \texttt{TSM2L} is \black{designed} to handle inputs with one tall-and-skinny matrix and one small regular-shaped matrix. Note that ``R'' or ``L'' means that the tall-and skinny matrix is multiplied on the right or left.

\subsection {Design of TSM2R}
In this section, we describe our proposed algorithm \texttt{TSM2R} for GEMM with a large regular-shaped matrix and a tall-and-skinny matrix.

\subsubsection{Insight on Tall-and-Skinny Input}
For regular-shaped GEMM ($m \times k$ matrix multiplies $k \times n$ matrix), the input matrices\black{' total} size is $O(mk+kn)$, while the computing time complexity is $O(mkn)$, so each element in \black{the} input matrices is used $O(m)$ or $O(n)$ times within the entire computation process. 
Since loading data \black{to the} GPU \black{from the} off-chip DRAM (i.e., global memory) to GPU is expensive and to avoid extensive data load operations, one common optimization for this kind of problem is minimizing the number of times each element needs to be loaded into the GPU by using fast on-chip memory (e.g., cache, registers) to enable data reuse. 
As the number of loads reduces, optimized GEMM tends to be compute-bound. 
For example, current GEMM implementation in cuBLAS library can reach near bare-metal performance on GPUs \cite{cublas-banchmark}.

However, unlike regular-shaped GEMM, when one matrix is tall-and-skinny (e.g., $n \ll m, k$), each element in the input matrices is used $O(n)$ times on average:
\begin{equation*}
    \resizebox{\width}{!}{
$\frac{(m \times k) \times n\ times + (k \times n) \times m\ times}{m\times k + k \times n} \approx O(n)$ times.}
\end{equation*}
Depending on the size of $n$ and target GPU peak computing power and memory throughput ratio, the computation can be either compute-bound or memory-bound. 
When $n$ gets smaller, the computation tends to be memory-bound. 
Otherwise, the problem tends to be compute-bound. 
In either case, the problem is always near the boundary between memory  bound  and compute  bound, so it is critical to design an algorithm that is optimized for both cases.

\subsubsection{Algorithm Design}
\black{Algorithm design plays a critical role in our proposed  optimizations.}
First, we need to consider how to fit the workload of our \texttt{TSM2R} into the programming model of CUDA (i.e., thread hierarchy). 
Although the workload can be easily decomposed into many independent smaller workloads, careful consideration \black{of the} workload distribution is still necessary, since any unnecessary performance penalty can cause \black{drastic} GPU resource under-utilization. Several factors are considered in our design:
\begin{enumerate}
\item Total number of global memory accesses;
\item \black{Shared and global memory access efficiency};
\item \black{Utilization of overall memory bandwidth};
\item Parallelism of overall workload;
\item On-chip memory utilization;
\item Streaming Multiprocessor (SM) utilization;
\item Optimization for compute \& memory-bound cases.
\end{enumerate}
To achieve good performance, there must exist enough active threads in each SM of \black{the} GPU to ensure proper instruction and memory access latency hiding. 
So, in our algorithm we divide the workload by assigning $n$ rows of matrix A to $n$ different threads. Each vector-matrix multiplication is assigned to one thread (i.e., ($A[i,:] \times B$)). 
The benefit is three-fold: 1) this ensures high parallelism and high SM occupancy;
2) since the number of elements of matrix A is much higher than matrix B, this kind of distribution ensures \black{that matrix A is accessed as little as possible}; 
3) it also enables high memory access efficiency and throughput, since all memory accesses to matrix A are naturally coalesced (assuming matrices are stored in column-major by convention). 

\setlength{\textfloatsep}{4pt}
\begin{algorithm}[t]
\small
\centering
\caption{\small Each thread's workload with inner product.}
\label{inner}
\begin{algorithmic}[1]
\REQUIRE{input matrix A ($m \times k$) and B ($k \times n$), output matrix C ($m \times n$)}
\FOR{$i=1$ to $n$}
	\FOR{$j=1$ to $k$}
		\STATE $C[global\_tid,i] += A[global\_tid,j] \times B[j,i]$
	\ENDFOR
\ENDFOR
\end{algorithmic}
\end{algorithm}

As for the vector-matrix multiplication assigned to each thread, to further reduce the number of memory accesses to matrix A, we use outer-product style computation instead of the \black{usual} inner-product style computation.  As shown in Algorithm \ref{inner}, if we use inner-product, each element of matrix A is repeatedly referenced $n$ times. On the other hand, if we use outer-product as shown in Algorithm \ref{outer}, each element of matrix A is referenced only once. (Please note, as we will discuss in later sections, when $n$ is larger than a certain threshold, elements in matrix A still need to be referenced more than once due to the limited resources available for each thread, but it is still far lower than using inner-product). \black{When matrix A is large}, the benefit is significant, since it greatly reduces the total number of global memory accesses during the entire GEMM computation. 
Also, the outer-product style does not bring any extra memory accesses to matrix B compared to inner-product style. The only cost for outer-product is extra registers holding $n$ intermediate results. However, with proper tuning, \black{the benefit of fewer memory accesses outweighs this cost}.  

\begin{algorithm}[t]
\small
\centering
\caption{\small Each thread's workload with outer product.}
\label{outer}
\begin{algorithmic}[1]
\REQUIRE{input matrix A ($m\times k$) and B ($k\times n$)}
\REQUIRE{output matrix C ($m \times n$)}
\STATE $Reg_{1:n} \leftarrow C[global\_tid,1:n]$
\FOR{$i=1$ to $k$}
	\STATE $Reg_{1:n} += A[global\_tid,i] \times B[i,1:n]$
\ENDFOR
\STATE $C[global\_tid,1:n]\leftarrow Reg_{1:n}$
\end{algorithmic}
\end{algorithm}

\subsubsection{Efficient Off-Chip Memory Access}
One key factor of optimizing memory intensive applications is ensuring high off-chip memory access efficiency. 
Depending on the GPU model type or runtime configurations, global memory (off-chip) accesses of threads within the same warp can \black{be} coalesced into 128 byte- or 32 byte-transactions \cite{cuda-doc} if their access addresses fall into the same 128 byte- or 32 byte-segments in global memory, which enables efficient use of memory bandwidth.  
Otherwise, the GPU still loads memory in 128 byte- or 32 byte-transactions, but it may contain unrequested data that are stored in neighbor addresses, which causes inefficient memory accesses. 

Since each thread reads one row of matrix A and the matrix is stored in column-major \black{format} by convention, memory accesses are naturally coalesced when threads within the same warp access elements on different rows but on the same column. 
So, 100\% memory access efficiency is achieved on matrix A. 
However, for matrix B, all threads access the same element at the same time, which results \black{in} a single memory transaction containing one requested element and several unrequested neighbor elements. 
So, only $\frac{8\ bytes}{128\ bytes} = 6.25\%$ or  $\frac{8\ bytes}{32\ bytes} = 25\%$ memory access efficiency is achieved for accessing 64-bit double floating point elements. 
Although the total number of elements in matrix B is small, given that each element needs to be accessed $n$ times, this inefficient access pattern can still greatly impact the overall performance.

To improve the efficiency of memory accesses to matrix B, we utilize shared memory in GPU. 
Since it is located on-chip, shared memory gives us the speed of L1 cache and it is fully programmable. 
Threads within one thread block can use shared memory to share data. So, one key advantage of shared memory is that it eliminates the need for the consistency between patterns of data loading and data using pattern, which enables us to load global memory in the most efficient way and keep the way that we use data as before. 

By using shared memory for accessing matrix B, we can reduce the total number of memory accesses and enable coalesced memory access. As shown in Algorithm \ref{shared}, for each iteration, instead of letting  threads request elements they need individually by themselves inefficiently, we now let a block of threads work together to fetch a tile of matrix B into the shared memory in a coalesce-compatible way (Line 11). Then during computation, each thread references elements in matrix B through the shared memory instead of loading each one of them individually from global memory. This reduces the total number of accesses to matrix B from global memory (from $n$ to $n/t_1$ per element). Also, threads in a same thread block fetch elements of matrix B column by column, which enables coalesced memory access and greatly improves memory-access efficiency to 100\%. Moreover, we also introduce three parameters: $t_1$, $t_2$, and $t_3$ in Algorithm \ref{shared}. These parameters are used for adjusting the performance and will be discussed in later sections.

\begin{algorithm}[t]
\small
\centering
\caption{\small \texttt{TSM2R} with shared memory.}
\label{shared}
\begin{algorithmic}[1]
\REQUIRE{input matrix A ($m \times k$) and B ($k \times n$), output matrix C ($m \times n$)}
\STATE $t_1 \leftarrow tile\_size\_B$, $t_2 \leftarrow tile\_size\_C$, $t_3 \leftarrow tile\_size\_A$
\STATE Register: $A_1,  A_2, ..., A_{t_3}$
\STATE Register: $C_1,  C_2, ... , C_{t_2}$
\STATE Shared Memory: currB with size $t_1 \times t_2$
\STATE Threads per thread block $\leftarrow t_1$
\STATE Total thread blocks $\leftarrow m / t_1$
\FOR{$p=1$ to $n$ with step size = $t_2$}
  \STATE $C_{1:t_2} \leftarrow C[global\_tid, p:p+t_2-1]$
  \FOR{$j=0$ to $k$ with step size = $t_1$}
      \STATEx \textcolor{blue}{\ \ \ \ \ \ \ \ \ \bf /* Load a tile of B into shared memory */}
      \STATE ThreadsSynchronization()
      \STATE \textcolor{blue}{$currB[global\_tid, 1:t_2]   \leftarrow B[j+global\_tid,p:p+t_2-1]$}
  	  \STATE ThreadsSynchronization()
      \FOR{$l=j$ to $j+t_1$ with step size = $t_3$}
      	  \STATEx \textcolor{red}{\ \ \ \ \ \ \ \ \ \ \ \ \bf /* Load a tile of A into registers*/}
          \STATE \textcolor{red}{$A_{1:t_3} \leftarrow A[global\_tid,l:l+t_3-1]$}
          \STATE $C_{1:t_2} +=  A_{1:t_3} \times currB[l:l+t_3,1:t_2]$
      \ENDFOR 
  \ENDFOR
  \STATE $ C[global\_tid ,p:p + t_2-1] \leftarrow C_{1:t_2}$
\ENDFOR
\end{algorithmic}
\end{algorithm}

\subsubsection{Optimizing Use of Shared Memory}
Although fast, elements in shared memory still need to be loaded into registers before using \black{them} \cite{ptx-doc}. 
Its \black{access} speed can affect the overall performance. 
Shared memory is divided into several same-sized memory banks for fast parallel accesses. Different threads \black{can access different memory banks} simultaneously. 
So, \black{having a} total \black{of} $b$ memory banks can speedup overall shared memory throughput by up to $b$ times compared to the throughput of one single memory bank. 
However, if $x$ threads in the same warp access different data from the same memory bank, \black{an} $x$-way bank conflict occur\black{s} and each request is processed sequentially, which dramatically reduces the accessing throughput by a factor of $1/x$.

In our algorithm, threads in the same thread block load data from global memory into shared memory column by column to enable fast coalesced global memory access. 
Then threads access data from shared memory row by row during computation. 
How we store elements in shared memory will affect how these elements are accessed from memory banks, which affects the throughput of shared memory. 
We have two ways of storing a tile of matrix B in shared memory: column-major storage and row-major storage. 
To choose between the two ways, we need to analyze and compare which way brings the least overall bank conflict. 
We assume the size of one tile of matrix B is $t_1 \times t_2$ and $t_1$ is the multiply of total number of memory banks $b$ for simplicity.

For column-major storage, elements (32-bit words or 64-bit words) in the same column of one tile of matrix B are stored in successive memory banks. 
So, for shared memory with $b$ memory banks, $t_1$ elements of one column are stored in $b$ different successive memory banks with each bank \black{storing} at most $\frac{t_1}{b}$ elements and \black{being} accessed by at most $\frac{warp\ size}{b}$ threads at the same time, which may potentially cause bank conflict if $\frac{warp\ size}{b}$ is greater than one.

\begin{figure}[t]
    \centering
%    \vspace*{-1em}
    \includegraphics[width=0.9\columnwidth]{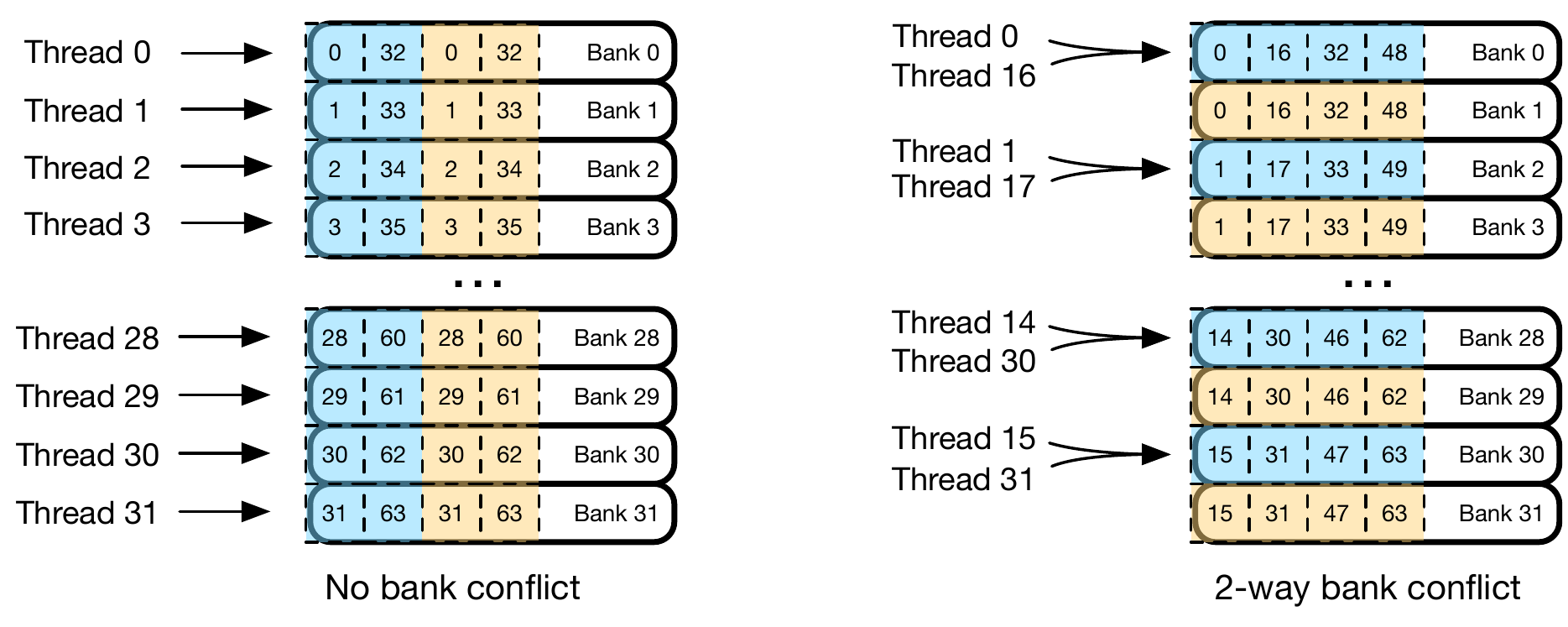}
    \vspace*{-0.5em}
    \caption{Comparing column-major (left) with row-major (right) storage for storing a $64 \times 2$ tile of matrix B in shared memory. Blue and yellow squares represent elements in the first and second column. When one warp of 32 threads accessing 32 elements in one column (e.g. element 0 to 31 of the first column), the column-major storage brings no bank conflict and row-major storage brings 2-way bank conflict, which reduces throughput by half.}
    \label{bank-conflict}
%    \vspace*{-0.5em}
\end{figure}

For row-major storage, elements in the same row of matrix B are stored in successive memory banks. 
So, elements of the same column are stored in $\frac{b}{t_2}$ different banks\black{, where} each bank stores $\frac{t_1 \times t_2}{b}$ elements from one column. Since each bank has $t_2$ times more elements from one column, each bank has at most $t_2$ times more thread\black{s} accessing \black{it} at the same time:  $\frac{warp\ size}{b} \times t_2$, which may also potentially cause bank conflict.

On modern NVIDIA GPUs, the $warp\ size$ is fixed to 32 and total number of banks is also 32 \cite{cuda-doc}, so column-major storage does not cause bank conflict, since each bank can only have up to one thread accessing. Row-major storage can cause up to $t_2$-way bank conflict, which decreases overall shared memory throughput to $\frac{1}{t_2}$ of the peak throughput. As shown in Figure \ref{bank-conflict}, we load a $64 \times 2$ matrix tile into shared memory using column-major storage (left) and row-major storage (right). 
When using column-major storage, threads in one warp all access different banks, so no bank conflict occurs. 
But when using row-major storage, 32 elements are stored in 16 banks causing 2-way bank conflict.
When accessing elements in shared memory for computation, threads in a warp all access the same element at \black{the same} time in our algorithm. Although multiple threads are accessing one bank, they are accessing the same element, so one broadcast is initiated, which does not cause bank conflict. It is the same for both storage styles. So, we choose column-major storage as it brings no bank conflict and potentially brings the highest shared memory throughput.

\subsubsection{Overlapping Computation and Memory Access Latency}

During execution, for each instruction issuing moment, each warp scheduler picks an eligible warp and send it to the corresponding component for execution. 
A warp becomes eligible only if all operands of its next instruction are ready. However, if a warp is loading data from global memory, it \black{will} take several hundred cycles before it can be ready for execution. 
To hide this long latency, we can either increase the number of threads \black{residing} in each SM to ensure there always exist eligible warps \cite{volkov2016understanding} or put independent instructions in between data loading and data consuming operations, so that warps are also eligible for execution during memory loading time. 
The first approach requires us to adjust the on-chip resource usage of each thread block. We will \black{save} that discussion \black{for} the next section. 
In this section, we aim to add independent instructions in between data loading and data consuming operations.

A shown in Algorithm \ref{shared}, Line 11 and 14 load data from global memory and Line 15 consumes data once data is loaded. 
However, due to data dependency, there is no independent instruction in between, so once each warp issues global memory access requests, it must wait for the requested elements to be ready before it can proceed to computation. 

So, to add independent instructions, we use data prefetching to mix data loading and \black{consumption} between neighbor iterations. 
Specifically, instead of letting each iteration loads data that is going to be used for current iteration, we let the data needed for current iteration to be loaded by the previous iteration, so that its calculation will not be blocked by data loading (since the data are ready). When doing calculation, it also loads data that is going to be used for the next iteration. By overlapping data loading and computation, we can significantly improve memory bandwidth and SM utilization. We apply data prefetching to both matrix A and B.

\begin{algorithm}
\footnotesize
\centering
\caption{\small \texttt{TSM2R} with shared memory and data prefetching.}
\label{prefetching}
\begin{algorithmic}[1]
\REQUIRE{input matrix A ($m \times k$) and B ($k \times n$), output matrix C ($m \times n$)}
\STATE $t_1 \leftarrow tile\_size\_B$, $t_2 \leftarrow tile\_size\_C$, $t_3 \leftarrow tile\_size\_A$
\STATE Register: $currA_1$, $currA_2$,...,$currA_{t_3}$
\STATE Register: $nextA_1$, $nextA_2$,...,$nextA_{t_3}$
\STATE Register: $nextB_1$, $nextB_2$,...,$nextB_{t_2}$
\STATE Register: $C_1$, $C_2$,...,$C_{t_2}$
\STATE Shared Memory: $currB$ with size $t_1 \times t_2$
\STATE Threads per thread block $\leftarrow t_1$
\STATE Total thread blocks $\leftarrow m / t_1$
\FOR{$p=1$ to $n$ with step size = $t_2$}
  
  \STATE $C_{1:t_2} \leftarrow C[global\_tid,p:p+t_2-1]$
  \STATEx {\ \ \ \ \ \bf /*load the first tile of A and B*/}
  \STATE \textcolor{blue}{$currB[local\_tid,1:t_2] \leftarrow B[local\_tid, p:p+t_2-1]$}
  \STATE \textcolor{red}{$currA_{1:t_3} \leftarrow A[global\_tid,1:t_3]$}
  \FOR{$j=0$ to $k$ with step size = $t_1$}
  \STATE ThreadsSynchronization()
  \STATEx \textcolor{blue}{\ \ \ \ \ \ \ \ \ \bf /*prefetch the next tile of B into registers*/}
      \IF {$j+t_1 < n$}
        \STATE \textcolor{blue}{$nextB_{1:t_2} \leftarrow B[j+t_1+local\_tid,p:p+t_2-1]$}
  	  \ENDIF
      \FOR{$l=j$ to $j+t_1$ with step size = $t_3$}
      \STATEx \textcolor{red}{\ \ \ \ \ \ \ \ \ \ \ \ \ \bf /*prefetch the next tile of A into registers*/}
      	  \IF {$l+t_3 < n$}
            \STATE \textcolor{red}{$nextA_{1:t_3} \leftarrow A[global\_tid,l+t_3:l+t_3+t_3-1]$}
          \ENDIF
          \STATE $C_{1:t_2} +=  currA_{1:t_3} \times currB[l:l+t_3,1:t_2]$
          \STATEx \textcolor{red}{\ \ \ \ \ \ \ \ \ \ \ \ \ \bf /*load the prefetched tile of A from nextA registers into currA registers*/}
          \STATE \textcolor{red}{$currA_{1:t_3} \leftarrow nextA_{1:t_3}$}
      \ENDFOR
      \STATE ThreadsSynchronization()
      \STATEx \textcolor{blue}{\ \ \ \ \ \ \ \ \ \bf /*load the prefetched tile of B from nextB registers to shared memory*/}
      \STATE \textcolor{blue}{$currB[local\_tid, 1:t_2] \leftarrow nextB_{1:t_2}$}
  \ENDFOR
  \STATE $ C[global\_tid,p:p+t_2-1] \leftarrow C_{1:t_2}$
\ENDFOR
\end{algorithmic}
\end{algorithm}

As shown in Algorithm \ref{prefetching}, we design our \texttt{TSM2R} with data prefetching. Note that \textit{global\_tid} and \textit{local\_tid} represent the (global) thread ID in the grid and the (local) thread ID in the block, respectively.
In Line 2 and 3, we allocate two sets of $t_3$ registers for storing current tile of elements of matrix A and next tile of element of matrix A for prefetching. In Line 4 and 6, we allocate $t_2$ registers for data prefetching of elements in matrix B, and allocate $t_1 \times t_2$ for storing currently loaded tile of matrix B. 
Note that we cannot store current tile of matrix B in registers, because elements in matrix B need to be shared between threads during computation.

\begin{figure}[]
% 	\vspace*{-1em}
    \centering
    \includegraphics[width=0.85\columnwidth]{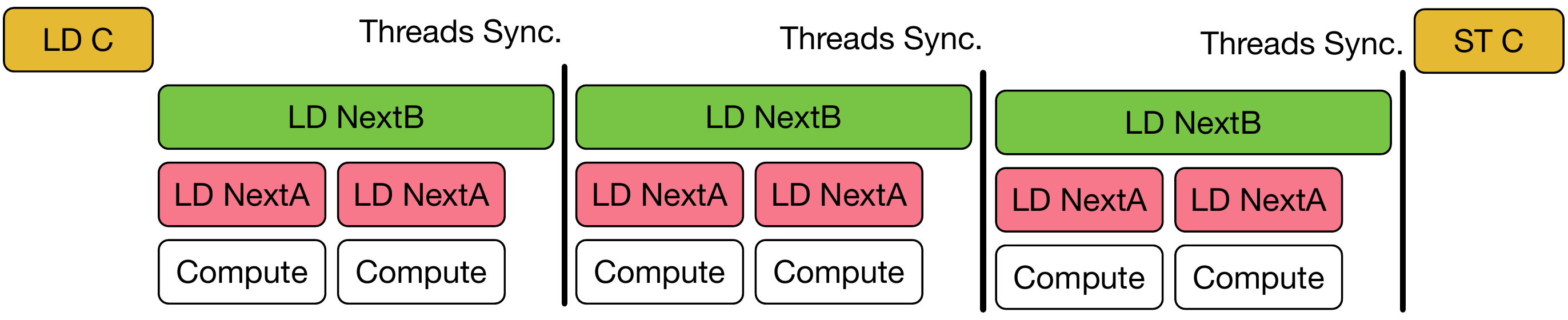}
    \vspace*{-0.5em}
    \caption{Example workload of one iteration of our optimized \texttt{TSM2R} with data prefetching.}
    \label{workload-prefetch}
%    \vspace{-1em}
\end{figure}

 \begin{figure}[]
    \centering
%    \vspace*{-1em}
    \includegraphics[width=0.9\columnwidth]{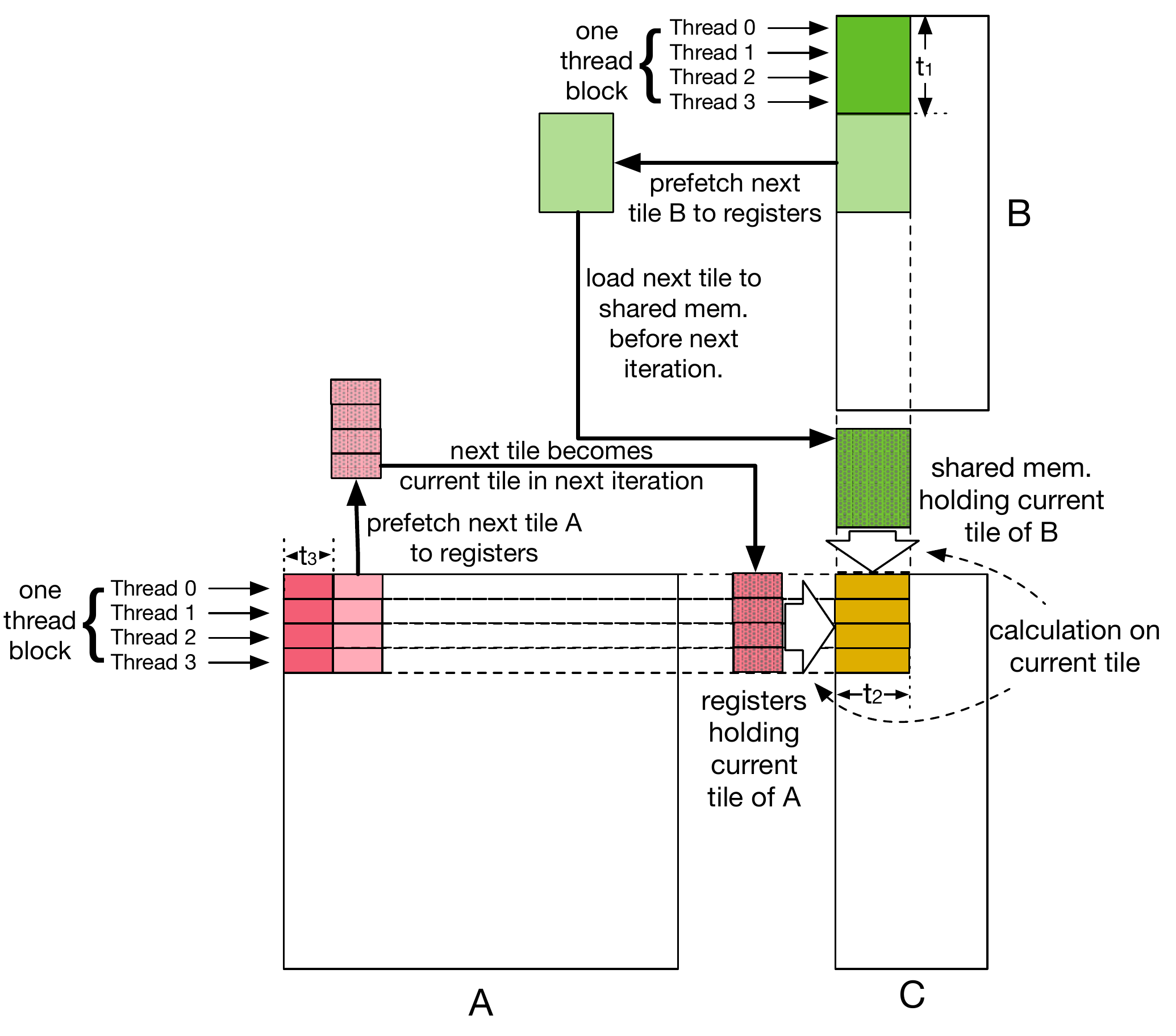}
    \vspace*{-0.5em}
    \caption{Matrix view of tall-and-skinny matrix matrix multiplication with data prefetching.}
    \label{workload-prefetch-matrix}
 %   \vspace*{-1em}
\end{figure}

Before the core computation iteration (Line 13-27), we pre-load current tile of matrix A and B into registers and shared memory (Line 11 and 12), so that computation can start immediately as soon as we enter the computation loop without being blocked by any data dependency. 
The main computation resides in Line 22. To overlap computation with memory accesses, we initiate loading for the next tile before the computation (Line 16 for matrix B and Line 20 for matrix A). 
We use two loops for loading matrix A and B, because we want to have the flexibility to adjust loading pace (tile size) differently for the two matrices. 
We will discuss this in the next subsection.
Figure\black{s} \ref{workload-prefetch} and \ref{workload-prefetch-matrix} show one iteration of our optimized \texttt{TSM2R} with data prefetching. \texttt{LD C} and \texttt{ST C} represent loading initial values from matrix C and storing final results back to matrix C. Each iteration we show three sub-iterations for loading matrix B. 
As we can see, we \black{compute and pre-load the next tile of matrix B concurrently} to improve memory bandwidth utilization. A \black{thread barrier} is inserted in the end of each iteration. 
For the \black{innermost} iteration, we do the actual computation and pre-load elements from matrix A each time. 
Please note that the length of each rectangle does not accurately represent the exact execution time length and the ratio between number of \texttt{LD nextA} and \texttt{LD nextB} is not necessarily two in actual computation. Also, we show one thread block with four threads only for illustration proposes.
As we will discuss in the next subsection that different parameter values can affect the length of each part and the ratio between number of \texttt{LD nextA} and \texttt{LD nextB}. Especially on the execution time of \texttt{LD nextA} and \texttt{Compute}, which will affect the characteristic of computation (i.e. memory-bound or compute-bound). 
Also, for simplicity, we ignore \black{the data movement from next tile to current tile that occurs in each iteration.}

\subsubsection{Parameters Definition}
In Algorithm \ref{shared} and Algorithm \ref{prefetching}, we introduced three adjustable parameters: $t_1$, $t_2$, and $t_3$. In this section, we first discuss how each parameter controls the computation of our \texttt{TSM2R}. 
Then, we introduce our performance model that estimates how certain performance metrics change with these parameters. 
Finally, we explain our strategies \black{for} choosing values for these parameters in order to achieve high GPU resource utilization and optimize overall performance.  
Please note that the following discussions are all based on Algorithm \ref{prefetching}.

\subsubsection{Behaviors of Parameters}
We first list the behaviors of each parameter below:
\begin{itemize}
\item $t_1$ specifies the number of rows of one tile of matrix B. To maximize use of available active threads and to avoid any inefficient thread execution caused by warp divergence, we let all threads in each thread block participat\black{e} in fetching elements of matrix B. For fast coalesced global memory access, we let each thread fetch one row, so $t_1$ is also the total number of threads in each thread block. Also, since we let \black{a} total \black{of} $m$ threads work on the computation, the total number of thread blocks can be calculated as: $m/t_1$.

\item $t_2$ specifies the number of elements in matrix C that each thread is working on at a time. It is used to divide the overall workload into several smaller workloads that are processed iteratively by each thread. \black{A smaller} workload makes each thread's SM resource usage smaller, which allows us to keep higher SM occupancy. However, dividing the workload means we need to load matrix A repeatedly for each small workload. So, there is a trade-off. $t_2$ also affects the ratio between total number of memory fetches and computation operations in core part of our algorithm, which allows us to adjust the computation to be compute or memory-bound (will be discussed later in detail).

\item $t_3$ specifies the number of elements in matrix A that each thread fetches at a time. Since elements fetches are independent to each other, they can be done without blocking each other, so $t_3$ can be used to adjust the memory loading concurrency.
\end{itemize}

\subsubsection{Performance Metrics Estimation}
In this section, we introduce our \black{parameter-}based performance model that is used to estimate three important performance metrics: SM occupancy, memory bandwidth utilization and computing power utilization. These estimations will be used for optimizing the overall performance.

\begin{itemize}
\item \textbf{Max SM occupancy estimation} \\
With these parameters we can calculate the max occupancy of each SM, which is defined as max number of active threads per SM. (Some works also use max number of warps, which is similar to ours. We found that using \black{the maximum number of threads} is more consistent across our performance models. We also choose \black{our} thread block size to be the dividend of this value to ensure \black{the} expected number of threads are active.)  This occupancy is mainly bound by the maximum hardware allowable number of threads ($HW\_MAX$) and on-chip memory utilization per thread. We first calculate the total number of registers utilized per thread. Since register utilization can potentially be optimized by the \texttt{nvcc} compiler, we use \black{the} maximum number of registers to estimate this value. First of all, there is a relatively fix amount of registers uses for CUDA initial setup, and we represent this amount as $C$. We get its amount through offline profiling. Then, we need two sets of $t_2$ registers for storing elements of matrix B for both next tile fetching and current tile calculation. Please note that although the current tile of matrix B is stored in shared memory, it still needs to be transferred to registers for calculation. Next, we need $t_2$ registers for keeping intermediate results of matrix C. Finally, we need two sets of $t_3$ registers for storing elements of matrix A for both next tile fetching and current tile calculation. So, the total number of registers is: 
\begin{equation*}
    \resizebox{\width}{!}{
$R_{thread} = (t_2 \times 3 + t_3 \times 2) \times \frac{bytes\_per\_element}{ bytes\_per\_register} + C$.}
\end{equation*}
As for shared memory, \black{although} it is allocated per thread block, we calculate the average amount of shared memory that each thread uses for consistent calculation here. Since the size of allocated shared memory per thread block is $t_1 \times t_2$, and as we will discuss earlier that we set $t_1 = threads\_per\_threadblock$, the amount of shared memory allocated for each thread on average is:
\begin{equation*}
    \resizebox{\width}{!}{
$S_{thread} = t_2 \times bytes\_per\_element$.}
\end{equation*}
So, the max SM occupancy can be calculated as:
\begin{equation*}
    \resizebox{\width}{!}{
$MaxOccup_{SM} = min(HW\_MAX, \frac{R_{SM}}{R_{thread}}, \frac{S_{SM}}{S_{thread}})$.}
\end{equation*}
In the above calculation, $R_{SM}$ and $S_{SM}$ stand for the max available registers and shared memory per SM. 
\item \textbf{Max memory bandwidth utilization estimation} \\
Next, we estimate the max memory bandwidth utilization of our algorithm when the computation is memory-bound. In this case, loading elements of matrix A dominates the computation instead of floating point calculations in our algorithm. So, we can estimate max memory bandwidth utilization using the maximum number of concurrent global memory accesses per SM. It can be calculated as:
\begin{equation*}
    \resizebox{\width}{!}{
$Concurrent_{mem} \approx MaxOccup_{SM} \times t_3$.}
\end{equation*}
Note \black{that} we only consider the memory accesses to matrix A here for simplicity. Since the majority \black{of} memory accesses are for matrix A, this only brings minor inaccuracy. Then, similar to \cite{volkov2016understanding, wong2010demystifying} we calculate the least number of concurrent memory accesses per SM needed to achieve max memory bandwidth utilization using Little's Law:
\begin{equation*}
    \resizebox{\width}{!}{
$Throughput_{max\_mem} = \frac{Peak\ Band.}{\#\_of\_SM \times core\_clock}$,}
\end{equation*}
\begin{equation*}
    \resizebox{\width}{!}{
$Concurrent_{max\_mem} = latency_{mem} \times Throughput_{max\_mem}$.}
\end{equation*}
The $latency_{mem}$ is the average global memory access latency, which is considered as a constant in our model and is obtained through offline profiling. The estimated memory bandwidth utilization is:
\begin{equation*}
    \resizebox{\width}{!}{
$Util_{mem} = \frac{Concurrent_{mem}}{Concurrent_{max\_mem}}.$
}
\end{equation*}

\item \textbf{Max computing power utilization estimation} \\
Next, we estimate the max computing power utilization of our algorithm when the computation is compute-bound. In this case, floating point calculation dominates the computation instead of memory accesses in our algorithm. So, we can estimate max computing power utilization using the maximum number of concurrent floating point operations per SM. It can be calculated as:
\begin{equation*}
    \resizebox{\width}{!}{
$Concurrent_{comp} = MaxOccup_{SM} \times t_3 \times t_2$.}
\end{equation*}
Then, also similar to \cite{volkov2016understanding} we calculate the least number of concurrent floating point operations per SM needed to achieve max computing power utilization using Little's Law:
\begin{equation*}
    \resizebox{\width}{!}{
$Throughput_{max\_comp} = \frac{Peak\ Perf.}{\#\_of\_SM \times core\_clock}$,}
\end{equation*}
\begin{equation*}
    \resizebox{\width}{!}{
$Concurrent_{max\_comp} = latency_{comp} \times Throughput_{max\_comp}$.}
\end{equation*}
The $latency_{comp}$ is the average latency of floating point operations in our calculations, which is considered as a constant in our model and is obtained through offline profiling. So, the estimated computing power utilization is:
\begin{equation*}
    \resizebox{\width}{!}{
$Util_{comp} = \frac{Concurrent_{comp}}{Concurrent_{max\_comp}}$.}
\end{equation*}
\item \textbf{Determine compute-bound or memory-bound} \\
Given parameters and GPU specification, we can determine whether the current computation is memory or compute-bound. This is mainly determined by the innermost loop (Line 20-24) of Algorithm \ref{prefetching}. The memory loading instructions (Line 21) \black{overlap the} computation (Line 23). Since Line 24 depends on memory loading results, it serves as an implicit synchronization point for memory load\black{ing} and computation. The time takes for the two parts will determine whether the current computation is compute-bound or memory-bound. So, we first estimate the time takes for computation and memory access as follows:
\begin{equation*}
    \resizebox{\width}{!}{
$time_{comp} = \frac{t_3\times t_2}{Peak\ Perf. \times \#\_of\_SM \times Occupancy_{SM}}$,}
\end{equation*}
\begin{equation*}
    \resizebox{\width}{!}{
$time_{mem} = \frac{t_3 \times bytes\_per\_elem.}{Peak\ Band. \times \#\_of\_SM \times Occupancy_{SM}}$.}
\end{equation*}
Then, by comparing the two time costs, we can determine whether the current computation is compute-bound or memory-bound.
\begin{equation*}
    \resizebox{\width}{!}{
$r = \frac{time_{comp}}{time_{mem}} = \frac{t_2}{bytes\_per\_elem.} \times \frac{Peak\ Band.}{Peak\ Perf.}$
}
\end{equation*}
As we can see, when $r$ is greater than one, the computation is compute-bound. Otherwise, the computation is memory-bound. Also, since we divide the original workload into several smaller workloads using $t_2$, this ratio is determined by $t_2$. By adjusting $t_2$, the actual computation can be shifted between compute and memory-bound. The boundary between the two cases can be calculated by setting the ratio $r = 1$, so we get a threshold for $t_2$: 
\begin{equation*}
    \resizebox{\width}{!}{
$t_2^{threshold} = \frac{Peak\ Perf.}{Peak\ Band.} \times bytes\_per\_elem.$
}
\end{equation*}

Similarly, we can also estimate \black{the} computation characteristic\black{s} of the original problem, in which the workload is not divided into smaller workloads. In this case, $t_2$ is always fixed to $k$.
So, by comparing $k$ with $t_2^{threshold}$ we can estimate the computation characteristic\black{s}. If $k$ is greater than $t_2^{threshold}$, the original problem is compute-bound; otherwise, it is memory-bound.
It can be easily seen, depending on the value of $t_2$ and $k$, the computation characteristic\black{s} of the current problem and original problem can be different, which can affect the overall performance. We discuss this in later part of this section.
\end{itemize}

\subsubsection{Deciding Parameters}
When choosing parameters, the first thing we should determine is whether we should optimize for computation or memory bandwidth. This is determined by whether the given \texttt{TSM2R} computation on the given GPU should be compute or memory-bound. In the last section, we proposed to estimate this characteristic by comparing $n$ and $t_2^{threshold}$, so that we can accordingly adjust parameters to optimize the computation.

In the case where original problem is memory-bound ($n \leq  t_2^{threshold}$), we need to keep the actual computation memory-bound also (let $1 \leq t_2 \leq n$) and optimize for memory bandwidth utilization. On the other hand, if the original problem is compute-bound ($n > t_2^{threshold}$), we first try to keep the actual computation compute-bound too (let $t_2^{threshold} \leq t_2 \leq n$) and optimize computing power utilization. However, in the case where $t_2^{threshold}$ is too high on the given GPU, we also try to optimize it for memory-bound (let $1 \leq t_2 \leq t_2^{threshold}$) and output the result parameters that deliver better performance.

Algorithm \ref{parameter-opt} shows the parameter optimization procedure for $t_2$ and $t_3$. We first determine the computation characteristic in Line 1. If it is memory-bound, we optimize for \black{the} total \black{global memory access time} (Line 4). Otherwise, we optimize for either total computation time (Line 9) or memory access time (Line 14).  Please note that we only count the total amount of memory accesses to matrix A for simplicity, since total accesses to matrix B is much less than matrix A, so this simplification only brings minor inaccuracy. Also, considering the total accesses to matrix B would bring one additional parameter ($t_1$), which can be hard to optimize since $t_1$ is also related to threads organization that is hard for modeling-based estimation. The memory bandwidth utilization term ($Util_{mem}$) and computing power utilization term ($Util_{comp}$) is calculated using the equation mentioned before. Since we have two parameters ($t_2$ and $t_3$) in our optimization target, we use Gradient Descent (GD) to do the optimization. In GD, based on our experience, we set initial value of both $t_2$ and $t_3$ to be 1, and step size to be 0.1. The stop threshold is set to be $10^{-4}$, since we do not need very accurate precision. The final $t_2$ and $t_3$ are rounded to the nearest integers.

\begin{algorithm}[t]
\footnotesize
\caption{\small Parameter optimization for \texttt{TSM2R}.}
\label{parameter-opt}
\begin{algorithmic}[1]
\IF{$n \leq  t_2^{threshold}$}
\STATE $Total\_memory \approx m \times k \times \frac{n}{t_2} \times bytes\_per\_elem.$
\STATE $Bandwidth = Peak Band. \times Util_{mem}$
\STATE Use Gradient Descent to Optimize ($t_2$ and $t_3$): $Time = \frac{Total\_memory}{Bandwidth }$ with $1 \leq t_2 \leq n$ and $1 \leq t_3$
\STATE \textbf{Output: $t_2$ and $t_3$ }
\ELSE
\STATE $Total\_flops = m \times k \times n \times 2$
\STATE $Compute\_power = Peak Perf. \times Util_{comp}$
\STATE Use Gradient Descent to Optimize ($t_2$ and $t_3$): $Time_1 = \frac{Total\_flops}{Compute\_power}$ with $t_2^{threshold} \leq t_2 \leq k$ and $1 \leq t_3$
\STATE $t_{2(time_1)} \leftarrow t_2$
\STATE $t_{3(time_1)} \leftarrow t_3$
\STATE $Total\_memory \approx n \times n \times \frac{k}{t_2} \times bytes\_per\_elem.$
\STATE $Bandwidth = Peak Band. \times Util_{mem}$
\STATE Use Gradient Descent to Optimize ($t_2$ and $t_3$) in $Time_2 = \frac{Total\_memory}{Bandwidth}$ with $1 \leq t_2 \leq t_2^{threshold}$ and $1 \leq t_3$
\STATE $t_{2(time_2)} \leftarrow t_2$
\STATE $t_{3(time_2)} \leftarrow t_3$
\IF{$Time_1 < Time_2$}
\STATE \textbf{Output: $t_{2(time_1)}$ and $t_{3(time_1)}$ }
\ELSE
\STATE \textbf{Output: $t_{2(time_2)}$ and $t_{3(time_2)}$ }
\ENDIF
\ENDIF
\end{algorithmic}
\end{algorithm}
%\vspace*{-0.5cm}

To optimize $t_1$, we \black{find} it only controls the number of threads in each thread block. Since the total number of threads is fixed to $m$, $t_1$ only determines how these threads are organized into thread blocks. There is trade-off: if $t_1$ is large, the total number of accesses to elements of matrix B is reduced, however, \black{a} large thread block means \black{a} large number of threads need to participate in the same synchronization, which may have \black{an} impact on performance. On the other hand, if $t_1$ is small,  the total number of accesses to elements of matrix B \black{is} higher, but the smaller thread block makes scheduling more flexible and efficient. It is hard to determine the optimum value of $t_1$ theoretically, so we use offline profiling to choose the best value. Specifically, once $t_2$ and $t_3$ are determined, we benchmark different $t_1$ values that can divide $MaxOccup_{SM}$ as mentioned earlier, and choose the $t_1$ for the best performance. Although $t_1$ seems to have direct effect on shared memory allocation (or max SM occupancy), it actually has limited impact on it, since we fix the amount of shared memory per thread ($S_{thread} = t_2 \times bytes\_per\_element$). 

\subsection{Design of TSM2L}

The algorithm proposed in the above sections---\texttt{TSM2R}---is optimized for the case where a large regular-shaped matrix multiples a tall-and-skinny matrix. In this section, we first propose a new algorithm \texttt{TSM2L} to handle the case where a tall-and-skinny matrix multiplies a small regular-shaped matrix. For example, an input matrix A of size $102400 \times 4$ multiples an input matrix B of size $4 \times 4$, where the tall-and-skinny matrix is multiplied on the left. We then introduce two different optimization approaches to overcome the bottleneck that this kind of tall-and-skinny input poses. 

\subsubsection{Performance Bottlenecks}

\begin{figure}[t]
    \centering
     \begin{subfigure}[t]{0.60\columnwidth}
        \includegraphics[width=\columnwidth]{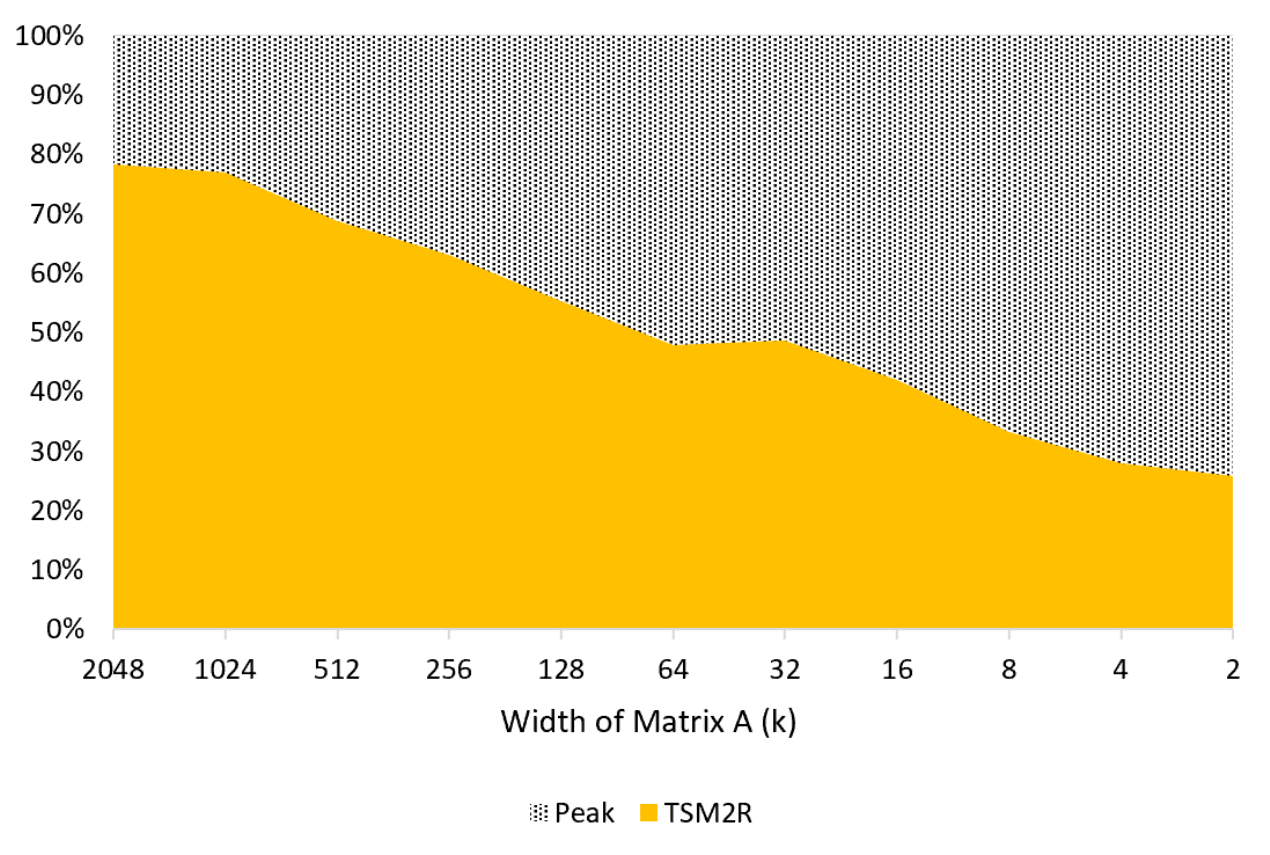}
    \end{subfigure}
    \vspace*{-0.5em}
    \caption{Memory bandwidth usage of very small values of k with double precision (m=15360, n=16).}
    \label{opt-motivation}
%    \vspace*{-0.5em}
\end{figure}

We start by adapting our previous algorithm \texttt{TSM2R} to handle the new case without further optimization. However, applying the algorithm to this case reveals a bottleneck. We evaluate \texttt{TSM2R} on this case with matrices of $15360 \times k$ and $k \times 16$---where $k$ varies---on an NVIDIA Tesla V100 GPU. As shown in Figure \ref{opt-motivation}, as the inner dimension $k$ decreases, the memory bandwidth usage also decreases.

To explain these results, we expand upon the performance model proposed in Section 3.6.2. This model assumes that the maximum theoretical occupancy is always achieved throughout the computation. However, on the one hand, since the algorithm loops $k \times \frac{n}{t2}$ times, and $k$ is very small, each thread does not perform enough workload to hide the latency, resulting in low occupancy. On the other hand, the program issues much fewer global memory reads than the case with large $k$, resulting in less efficient memory usage. Therefore, \texttt{TSM2R} performs in a \textbf{latency-bound mode} (neither compute-bound nor memory-bound) on this input case (i.e., a tall-and-skinny matrix multiplying a small matrix), as indicated in a prior study \cite{volkov2016understanding}.

\subsubsection{Proposed Optimizations}

\begin{algorithm}[t]
\small
\caption{Proposed optimization-1 for \texttt{TSM2L}.}
\label{alg-opt1}
\begin{algorithmic}[1]
\REQUIRE{input matrix A ($m \times k$) and B ($k \times n$)}
\REQUIRE{output matrix C ($m \times n$)}
\STATE $tcf \leftarrow thread\_count\_factor$
\STATE $t_1 \leftarrow tile\_size\_B$, $t_2 \leftarrow tile\_size\_C$, $t_3 \leftarrow tile\_size\_A$
\STATE Register: $currA_1$, $currA_2$,...,$currA_{t_3}$
\STATE Register: $nextA_1$, $nextA_2$,...,$nextA_{t_3}$
\STATE Register: $nextB_1$, $nextB_2$,...,$nextB_{t_2}$
\STATE Register: $C_1$, $C_2$,...,$C_{t_2}$
\STATE Shared Memory: $currB$ with size $t_1 \times t_2$
\STATE Threads per thread block $\leftarrow t_1$
\STATE Total thread blocks $\leftarrow m /(t_1 \times tcf)$
\STATE Total threads $\leftarrow m / tcf$

\STATEx { \bf /*loop over all the horizontal tiles of matrix A*/}
\FOR{$r=global\_tid$ to $m$ with step size = Total threads}
\STATE \textcolor{red}{Perform Line 9-29 of Algorithm \ref{prefetching} with all occurrences of the identifier $global\_tid$ replaced by the identifier $r$}
\ENDFOR
\end{algorithmic}
\end{algorithm}

Based on these observations, we design two \black{further} optimizations for \texttt{TSM2L}. Both optimizations intend to trade warp latency for memory-access latency by launching fewer threads. As a result, each thread performs more computation, and the accumulated warp latency can be replaced by the memory-access latency. Since we are launching fewer threads than the number of rows of matrix $A$, we must divide it into several horizontal tiles. Here we introduce a new parameter $tcf$ to represent the tile number of matrix $A$ in our algorithm. We launch $\frac{m}{tcf}$ threads in the new kernel.

The first optimization involves dividing the multiplication into $tcf$ parts, where each part consists of multiplying a $\frac{m}{tcf}$-row tile of matrix $A$ by the entire matrix $B$. In essence, this optimization repeats the \texttt{TSM2R} algorithm once for each tile of matrix $A$. In this optimization, each element of matrix $A$ is still accessed $\frac{n}{t_2}$ times, though each element of $B$ is loaded $tcf \times \frac{n}{t_2}$ times, which is $tcf$ times more than that in \texttt{TSM2R}. We describe the detail in Algorithm \ref{alg-opt1}. 

\begin{algorithm}[]
\scriptsize
\caption{Proposed optimization-2 for \texttt{TSM2L}.}
\label{alg-opt2}
\begin{algorithmic}[1]
\REQUIRE{input matrix A ($m \times k$) and B ($k \times n$), output matrix C ($m \times n$)}
\STATE $tcf \leftarrow thread\_count\_factor$
\STATE $t_1 \leftarrow tile\_size\_B$, $t_2 \leftarrow tile\_size\_C$, $t_3 \leftarrow tile\_size\_A$
\STATE Register: $currA_1$, $currA_2$,...,$currA_{t_3}$
\STATE Register: $nextA_1$, $nextA_2$,...,$nextA_{t_3}$
\STATE Register: $nextB_1$, $nextB_2$,...,$nextB_{t_2}$
\STATE Register: $C_1$, $C_2$,...,$C_{t_2}$
\STATE Register: $nextC_1$, $nextC_2$,...,$nextC_{t_2}$
\STATE Shared Memory: $currB$ with size $t_1 \times t_2$
\STATE Threads per thread block $\leftarrow t_1$
\STATE Total thread blocks $\leftarrow m /(t_1 \times tcf)$
\STATE Total threads $\leftarrow m / tcf$

\FOR{$p=1$ to $n$ with step size = $t_2$}
  \STATE \textcolor{blue}{$currB[local\_tid,1:t_2] \leftarrow B[local\_tid, p:p+t_2-1]$}
  \STATE \textcolor{red}{$currA_{1:t_3} \leftarrow A[global\_tid,1:t_3]$}
  %\STATE\Comment{Load $currB$ and $currA$ as described in Line 12 and 13 of \textbf{Algorithm \ref{prefetching}}}
  \FOR{$j=0$ to $k$ with step size = $t_1$}
  \STATE ThreadsSynchronization()
  \STATEx \textcolor{blue}{\ \ \ \ \ \ \ \ \ \bf /*prefetch the next tile of B into $nextB$*/}
  \STATE \textcolor{blue}{$nextB_{1:t_2} \leftarrow B[j + t_1 + local\_tid - 1,p:p+t_2-1]$}
  \STATE $C_{1:t_2} \leftarrow C[global\_tid,p:p+t_2-1]$
    \STATEx {\ \ \ \ \ \ \ \ \ \bf /*loop over all the horizontal tiles of matrix A*/}
    \FOR{$r=global\_tid$ to $m$ with step size = Total threads}
      \STATEx \textcolor{blue}{\ \ \ \ \ \ \ \ \ \ \ \ \ \bf /*prefetch the next tile of C into registers*/}
      \IF {$r$ $+$ Total threads $\leq m$}
        \STATE $nextC_{1:t_2} \leftarrow C[r + $Total threads$, p:p + t_2 - 1]$
      \ENDIF
      \STATE \textcolor{red}{Perform Line 18-25 of Algorithm \ref{prefetching}, with all occurrences of the identifier $global\_tid$ replaced by the identifier $r$}
      \STATEx \textcolor{blue}{\ \ \ \ \ \ \ \ \ \ \ \ \ \bf /*store the sum so far in $C$*/}
      \STATE $C[r, p:p + t_2 - 1] \leftarrow C_{1:t_2}$
      \STATEx \textcolor{blue}{\ \ \ \ \ \ \ \ \ \ \ \ \ \bf /*load the prefetched tile C from nextC to C*/}
      \STATE $C_{1:t_2} \leftarrow nextC_{1:t_2}$
    \ENDFOR
    \STATE ThreadsSynchronization()
    \STATEx{\textcolor{blue}{\ \ \ \ \ \ \ \ \ \bf /*load the prefetched tile B from nextB to shared memory*/}}
    \STATE \textcolor{blue}{$currB[local\_tid, 1:t_2] \leftarrow nextB_{1:t_2}$}
  \ENDFOR
\ENDFOR
\end{algorithmic}
\end{algorithm}

The second optimization is to interleave the computation of the tiles, rapidly loading elements from different tiles of matrix $A$ and loading and storing intermediate sums in matrix $C$. Once a $t1 \times t2$ tile of matrix $B$ is loaded, the intermediate results are loaded, computed, and stored for each tile of matrix $A$. The $C$ register set is loaded with values from matrix $C$ which contains the product accumulated so far. After the computation is finished, the values are stored to matrix $C$ again as the next tile of matrix $A$ is prepared for computation. To quickly switch between tiles, values from matrix $C$ are prefetched in addition to the prefetching already described in Algorithm \ref{prefetching}. A new set of registers, $nextC_{1:t_2}$, is used to store the values of $C$ associated with the next tile of $A$. The elements of $A$ and $B$ are accessed only $\frac{n}{t_2}$ times, though each element of $C$ is accessed $\frac{k}{t_1} \times \frac{n}{t_2}$ times. However, since we do not achieve either high occupancy or high memory bandwidth in this case, we are not as concerned about issuing more memory read instructions. We describe the detail in Algorithm \ref{alg-opt2}.

\begin{figure}[t]
    \centering
    \begin{subfigure}[t]{0.49\columnwidth}
        \includegraphics[width=\columnwidth]{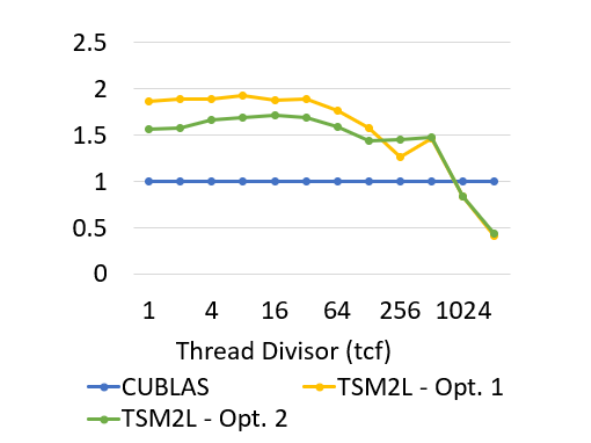}
        \caption{Speedup (single)}
    \end{subfigure}
    \begin{subfigure}[t]{0.49\columnwidth}
        \includegraphics[width=\columnwidth]{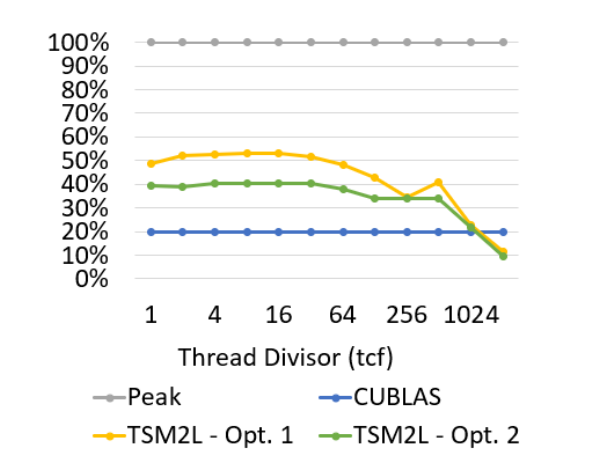}
        \caption{Mem. bw. util. (single)}
    \end{subfigure}
    \begin{subfigure}[t]{0.49\columnwidth}
        \includegraphics[width=\columnwidth]{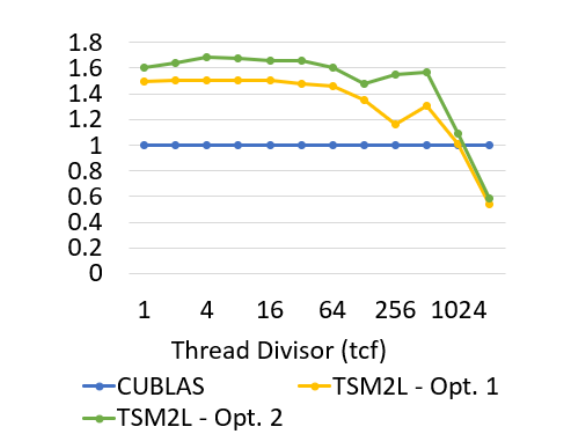}
        \caption{Speedup (double)}
    \end{subfigure}
    \begin{subfigure}[t]{0.49\columnwidth}
        \includegraphics[width=\columnwidth]{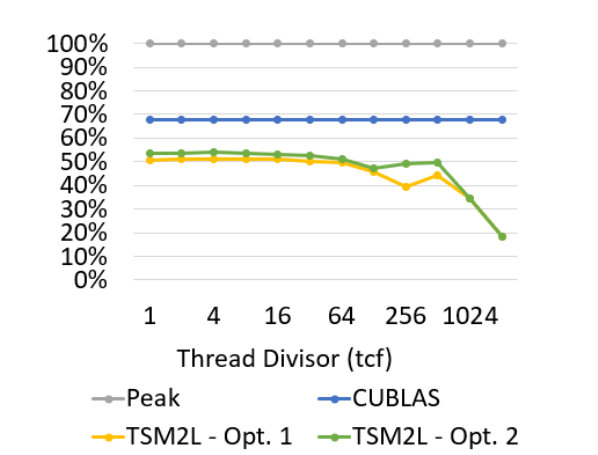}
        \caption{Mem. bw. util. (double)}
    \end{subfigure}
    \vspace*{-0.5em}
    \caption{Performance comparison with single and double precision using different $tcf$ (m=$10^7$, k=n=16).}
    \label{opt-examination}
 %   \vspace*{-1.5em}
\end{figure}

Figure \ref{opt-examination} illustrates the effects of the two optimizations on both performance and memory bandwidth usage. As fewer and fewer threads are launched, the impact of warp latency is replaced with that of different kinds of latency such as memory bandwidth latency. As a result, computation time decreases and memory access bandwidth increases in this case. Note that for this case the number of threads launched must be reduced to at least $\frac{1}{64}$ of $m$ before a significant decline in speedup or memory bandwidth utilization occurs. This is \black{because} the kernel must manage so many threads that \black{perform} little work when $m = 10^7$. 

Therefore, we must choose an appropriate $tcf$, determining the number of threads to launch for each kernel. If the algorithm is launched with an insufficient number of threads, the parallelism becomes too low and hence the performance would suffer. If the algorithm is launched with too many threads, the performance would be impacted by warp latency just as it does in the naive adaptation of \texttt{TSM2R}. We thus must determine an appropriate $tcf$ for each target system with offline profiling.

\begin{table*}[ht]
\sffamily
\footnotesize
\centering
% Please forgive my abysmal table formatting
\begin{tabular}{@{}rl@{}}
    % \midrule
\toprule
      \textbf{\texttt{TSM2R}}
    & Large square or rectangular matrix by tall-and-skinny matrix
    \\
      \textbf{\texttt{TSM2L}}
    & Tall-and-skinny matrix by small square matrix
    \\
    \midrule
      \textbf{\texttt{TSM2R} optimizations}
    & \textbf{Compute and memory-bound cases}
    \\
      Algorithm \ref{inner}
    & Inner product only
    \\
      Algorithm \ref{outer}
    & Outer product: saves global memory accesses
    \\
      Algorithm \ref{shared}
    & Shared memory: more efficient global accesses to matrix B
    \\
      Algorithm \ref{prefetching}
    & Data prefetch: overlap compute and memory operations 
    \\
    \midrule
      \textbf{\texttt{TSM2L} optimizations}
    & \textbf{Latency bound cases}
    \\
      Algorithm \ref{alg-opt1}
    & Divide matrix A into horizontal tiles: compute each tile sequentially
    \\
      Algorithm \ref{alg-opt2}
    & Divide matrix A into horizontal tiles: interleave the computation of each tile
    \\
    \midrule
      \textbf{Performance Model}
    &
    \\
      Parameter $t_1$
    & Number of rows of a tile of Matrix B
    \\
      Parameter $t_2$
    & Number of elements of C each thread computes at a time
    \\
      Parameter $t_3$
    & Number of elements of A each thread fetches at a time
    \\
      \textbf{$Util_{comp}$}
    & Computing power utilization term
    \\
      \textbf{$Util_{mem}$}
    & GPU memory bandwidth utilization term
    \\
      \textbf{$t_2^{threshold}$}
    & $= \frac{Peak\ Perf.}{Peak\ Band.} \times bytes\_per\_elem.$
    \\
    & Determines whether a computation is compute-bound or memory-bound
    \\
    \bottomrule
\end{tabular}%
\caption{{Summary of \texttt{TSM2R} and \texttt{TSM2L} design.}}
\label{designsummary}
\end{table*}

\subsection{Design Summary}
\black{We summarize the design of \texttt{TSM2R} and \texttt{TSM2L}, including our performance model, in Table \ref{designsummary}.}

\section{Experimental Evaluation}
\label{experiments}

\subsection{Experiments Setup}

\begin{table*}[ht]
\sffamily
\scriptsize
\centering

\begin{tabular}{@{} lrrrrr@{}}

% \toprule
    &
    &
    & \textbf{Darwin}
    & 
    & \textbf{PantaRhei}
    \\ 
    % \midrule
    \toprule
    \textbf{CPU}
    &
    &
    \multicolumn{2}{r}{Intel Xeon E5-2650v2}
    &
    & Intel Xeon Gold 6148
    \\
    \textbf{Memory}
    &
    & 
    & 251 GB
    & 
    & 384 GB
    \\
    \textbf{GPU}
    &
    & Tesla K40c
    & Tesla M40
    & Tesla P100
    & Tesla V100
    \\
    \textbf{Architecture}
    &
    & Kepler
    & Maxwell
    & Pascal
    & Volta
    \\
    \textbf{GPU Memory}
    &
    & 12 GB
    & 12 GB
    & 16 GB
    & 16 GB
    \\
    \textbf{Peak Performance (Single)}
    &
    & 5046 GFLOPS
    & 6844 GFLOPS
    & 10600 GFLOPS
    & 15000 GFLOPS
    \\
    \textbf{Peak Performance (Double)}
    &
    & 1430 GFLOPS
    & 213 GFLOPS
    & 4600 GFLOPS
    & 7500 GFLOPS
    \\
    \textbf{Memory Bandwidth}
    &
    & 288 GB/s
    & 288 GB/s
    & 720 GB/s
    & 900 GB/s
    \\
    % \bottomrule
\end{tabular}
\caption{Experimental platforms with detailed GPU information.}
\vspace{-2mm}
\label{cluster}
\end{table*}

We implement our \texttt{TSM2R} and \texttt{TSM2L} using CUDA C for single and double \black{precision} floating-point input. We disable compiler auto unrolling \black{in favor of explicit loop unrolling} for better control on register allocation. \black{Note that since our proposed algorithms mainly target traditional scientific computing applications rather than machine learning applications, we omit an evaluation on half-precision input.
%Since our algorithms' input configurations have primary applications in scientific rather than machine learning, we omit an analysis of half-precision input.
}
We evaluate our optimized implementations on two heterogeneous testbed clusters, which are Darwin \cite{darwin} at Los Alamos National Laboratory and PantaRhei \cite{pantarhei} at the University of Alabama.
We run each test on a single GPU node with single GPU. We conduct our tests on four different commonly used modern NVIDIA GPUs with four different micro-architectures: Kepler, Maxwell, Pascal, and Volta. 
For Kepler GPU, we use Tesla K40c, which has 1430 GFLOPS peak double floating-point performance and 288 GB/s memory bandwidth. 
For Maxwell GPU, we use Tesla M40, which has 213 GFLOPS peak double floating-point performance and 288 GB/s memory bandwidth. 
For Pascal GPU, we use Tesla P100, which has 4600 GFLOPS peak double floating-point performance and 720 GB/s memory bandwidth. 
For Volta GPU, we use Tesla V100, which as 7500 GFLOPS peak double floating-point performance and 900 GB/s memory bandwidth. \black{We provide more information about our experimental clusters and GPUs in Table \ref{cluster}.}

For comparison, we compare our \texttt{TSM2R} and \texttt{TSM2L} with GEMM in the current latest cuBLAS library \cite{cublas} and latest BLASX library \cite{wang2016blasx}. Also, we try to compare our work with KBLAS \cite{abdelfattah2016kblas}, however since its GEMM kernel is based on cuBLAS, its performance is identical to cuBLAS, so we omitted its results. Each test is repeated multiple times to reduce noise and timed using CUDA Events API. We measure performance by calculating the performance of \texttt{FAMD} instructions. We also measure the global memory throughput using \texttt{nvprof} on the command line with \texttt{--metrics gld\_throughput} option. In addition, we use \texttt{--metrics gld\_efficiency} option to verify if 100\% global memory access efficiency is achieved in our optimization.% during development.

Our input matrices are initialized with random floating-point numbers between 0 and 1. We test the multiplication of a large squared matrix and a tall-and-skinny matrix for \texttt{TSM2R} and the multiplication of a tall-and-skinny matrix and a small squared matrix for \texttt{TSM2L}.
Specifically, for \texttt{TSM2R}, the size of the large regular-shaped matrix is from $10240 \times 10240$ to $30720 \times 30720$, and the size of the tall-and-skinny matrix ranges from $10240 \times n$ to $30730 \times n$ with $n$ equals 2, 4, 8, and 16.
For \texttt{TSM2L}, the size of the tall-and-skinny matrix ranges from $10^4 \times k$ to $10^7 \times k$ with $k$ equals 8 or 16, and the size of the small regular-shaped matrix is 8 or 16.

\subsection{Evaluation of \texttt{TSM2R}}
In this section, we first evaluate the performance of \texttt{TSM2R} with different input sizes and compare it with state-of-the-art libraries on K40c, M40, P100, and V100.

\subsubsection{Tests with Different Optimization Combinations}
We use the GEMM in cuBLAS as \black{our} comparison baseline. We apply different combinations of optimization in \texttt{TSM2R} and compare them with GEMM in cuBLAS and BLASX. We have \black{in total} four versions of \texttt{TSM2R}:

\begin{itemize}
\item \texttt{V0}: the most straightforward inner product version as described in Algorithm \ref{inner};
\item \texttt{V1}: the outer version as in Algorithm \ref{outer}. This version reduces the total number of global memory accesses from algorithm level;
\item \texttt{V2}: based on outer production version as in Algorithm \ref{outer}, we add the use of shared memory, leading to more efficient global memory access to matrix B;
\item \texttt{V3}: based on the outer production version as in Algorithm \ref{outer} and the use of shared memory, we add data prefetch. This is the best version of our optimized implementation, which is described in Algorithm \ref{prefetching}.
\end{itemize}

\begin{figure}[t!]
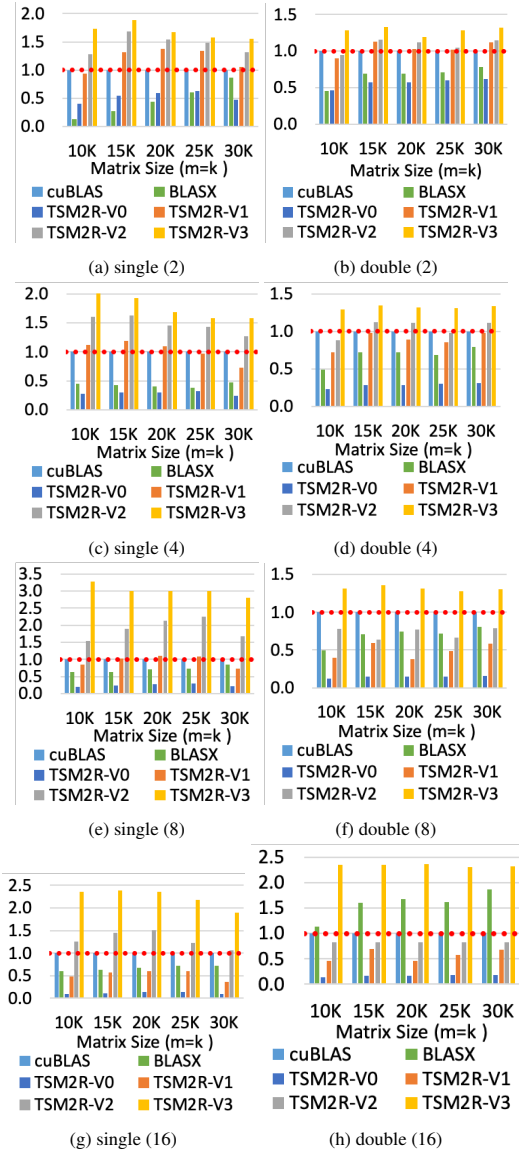

    \centering
    \begin{subfigure}[t]{0.36\columnwidth}
        \includegraphics[width=\columnwidth]{figures/k40-2s-speedup.pdf}
        \caption{single (2)}
    \end{subfigure}
    \begin{subfigure}[t]{0.36\columnwidth}
        \includegraphics[width=\columnwidth]{figures/k40-2-speedup.pdf}
        \caption{double (2)}
    \end{subfigure}
    \begin{subfigure}[t]{0.36\columnwidth}
        \includegraphics[width=\columnwidth]{figures/k40-4s-speedup.pdf}
        \caption{single (4)}
    \end{subfigure}
    \begin{subfigure}[t]{0.36\columnwidth}
        \includegraphics[width=\columnwidth]{figures/k40-4-speedup.pdf}
        \caption{double (4)}
    \end{subfigure}
    \begin{subfigure}[t]{0.36\columnwidth}
        \includegraphics[width=\columnwidth]{figures/k40-8s-speedup.pdf}
        \caption{single (8)}
    \end{subfigure}
    \begin{subfigure}[t]{0.36\columnwidth}
        \includegraphics[width=\columnwidth]{figures/k40-8-speedup.pdf}
        \caption{double (8)}
    \end{subfigure}
    \begin{subfigure}[t]{0.36\columnwidth}
        \includegraphics[width=\columnwidth]{figures/k40-16s-speedup.pdf}
        \caption{single (16)}
    \end{subfigure}
    \begin{subfigure}[t]{0.4\columnwidth}
        \includegraphics[width=\columnwidth]{figures/k40-16-speedup.pdf}
        \caption{double (16)}
    \end{subfigure}
%    \vspace*{-0.5em}
    \caption{Speedup comparison with single and double precision on K40c (n = 2, 4, 8, 16).}
    \label{k40-speedup}
%    \vspace*{-0.5em}
\end{figure}

\black{We provide detailed performance breakdowns on K40c and V100, but our} optimization behaves similarly on other GPUs. To evaluate our optimization, we need to determine by which resource our program is bounded. Since, $t_{2(k40c)}^{threshold} \approx 40$, the computation is always memory bound for the given $n$ values. The optimized parameters are: $t_2 = n$, $t_3 = 4$, and $t_1 = 128$. The parameters are only applied to the last to versions of \texttt{TSM2R}. Figure \ref{k40-speedup} shows the speedup of different versions in single and double precision. From the results, we can see that the \texttt{TSM2R-V0} suffers from poor performance due to the requirement of much higher number of global memory accesses in the inner product version. \texttt{TSM2R-V1}, on the other hand, significantly improves the performance compared to \texttt{TSM2R-V0} (2.2x$\sim$4.7x faster), since it requires much lower number of global memory accesses. \texttt{TSM2R-V2} further improves the efficiency of global memory access to matrix B, which plays a vital role in the overall performance. In addition, the shared memory shares tiles of matrix B between threads within a thread block also reduced the total number of memory accesses to matrix B. This leads to additional 1.1x to 2.1x speedup. Finally, the data prefetch introduced in \texttt{TSM2R-V3} further mitigate\black{s} the memory access bottleneck, which brings additional 1.3x$\sim$3.5x speedup \black{(1.9x on average)}.

\subsubsection{Memory Throughput Analysis}

\begin{figure}[t!]
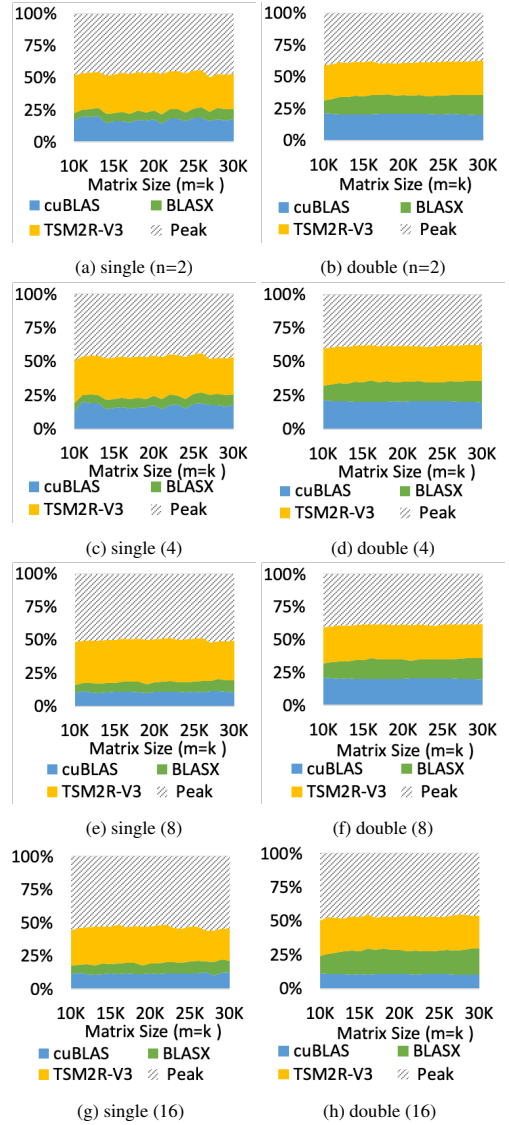

    \centering
    \begin{subfigure}[t]{0.36\columnwidth}
        \includegraphics[width=\columnwidth]{figures/k40-2s-mem.pdf}
        \caption{single (n=2)}
    \end{subfigure}
    \begin{subfigure}[t]{0.36\columnwidth}
        \includegraphics[width=\columnwidth]{figures/k40-2-mem.pdf}
        \caption{double (n=2)}
    \end{subfigure}
    \begin{subfigure}[t]{0.36\columnwidth}
        \includegraphics[width=\columnwidth]{figures/k40-4s-mem.pdf}
        \caption{single (4)}
    \end{subfigure}
    \begin{subfigure}[t]{0.36\columnwidth}
        \includegraphics[width=\columnwidth]{figures/k40-4-mem.pdf}
        \caption{double (4)}
    \end{subfigure}
    \begin{subfigure}[t]{0.36\columnwidth}
        \includegraphics[width=\columnwidth]{figures/k40-8s-mem.pdf}
        \caption{single (8)}
    \end{subfigure}
    \begin{subfigure}[t]{0.36\columnwidth}
        \includegraphics[width=\columnwidth]{figures/k40-8-mem.pdf}
        \caption{double (8)}
    \end{subfigure}
    \begin{subfigure}[t]{0.36\columnwidth}
        \includegraphics[width=\columnwidth]{figures/k40-16s-mem.pdf}
        \caption{single (16)}
    \end{subfigure}
    \begin{subfigure}[t]{0.36\columnwidth}
        \includegraphics[width=\columnwidth]{figures/k40-16-mem.pdf}
        \caption{double (16)}
    \end{subfigure}
    \vspace*{-0.5em}
    \caption{Memory bandwidth utilization comparison on K40c (n = 2, 4, 8, 16).}
    \label{k40-mem}
%    \vspace*{-1.5em}
\end{figure}

Figure \ref{k40-mem} shows the memory throughput of \texttt{TSM2R-V3}, cuBLAS and BLASX in both single and double precision on K40c GPU. Result show that \texttt{TSM2R} brings 12.5\%$\sim$26.6\% (17.6\% on average) improvement on memory bandwidth utilization compared with cuBLAS and 20.1\%$\sim$38.8\% (24.3\% on average) improvement compared with BLASX.

\begin{figure}
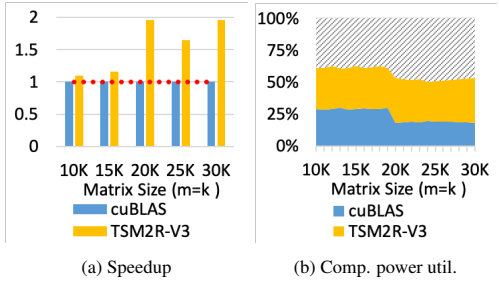
%[t]
    \centering
    \begin{subfigure}[t]{0.36\columnwidth}
        \includegraphics[width=\columnwidth]{figures/m40-16-speedup.pdf}
        \caption{Speedup}
    \end{subfigure}
    \begin{subfigure}[t]{0.36\columnwidth}
        \includegraphics[width=\columnwidth]{figures/m40-16-comp.pdf}
        \caption{Comp. power util.}
    \end{subfigure}
    \vspace*{-0.5em}
    \caption{Speedup and computing power utilization comparison with double precision on M40 (n = 16).}
    \label{m40-16}
    \vspace*{-1em}
\end{figure}

\subsubsection{Tests on Different Micro-architectures}

\begin{figure}[t]
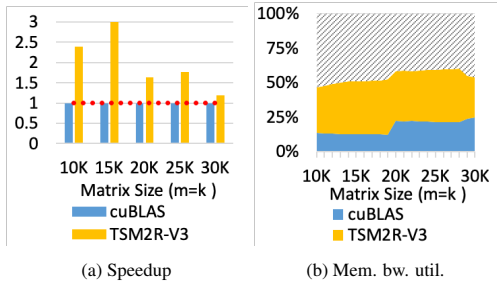

    \centering
    \begin{subfigure}[t]{0.36\columnwidth}
        \includegraphics[width=\textwidth]{figures/p100-16-speedup.pdf}
        \caption{Speedup}
    \end{subfigure}
    \begin{subfigure}[t]{0.36\columnwidth}
        \includegraphics[width=\columnwidth]{figures/p100-16-mem.pdf}
        \caption{Mem. bw. util.}
    \end{subfigure}
    \vspace*{-0.5em}
    \caption{Speedup and memory bandwidth utilization comparison with double precision on P100 (n = 16).}
    \label{p100-16}
 %   \vspace*{-1.5em}
\end{figure}

In addition to Kepler micro-architecture, we also conduct \black{tests} on newer Maxwell, Pascal, and Volta GPUs. Similar \black{to the} Kepler GPU, we get $t_{2(m40)}^{threshold} \approx 6$ and $t_{2(p100)}^{threshold} \approx 50$. Tesla M40 has \black{lower} computing power, so the computation with input with $n = 16$ is compute bound. Our parameter optimization procedure also output parameters in favor of computing optimization: $t_2 = 8$, $t_3 = 4$, and $t_1 = 256$. As shown in Figure \ref{m40-16}, our optimized implementation achieves 1.1x -1.9x (1.47x on average) speedup on Tesla M40 with 7\% to 37.3\% (20.5\% on average) computing power usage improvement compared to the GEMM function in cuBLAS 9.0. P100 has much stronger computing power, as we can see the computation with input with $n = 16$ is memory bound. Our parameter optimization procedure also output parameters in favor of memory optimization: $t_2 = 4$, $t_3 = 4$, and $t_1 = 128$. As shown in Figure \ref{p100-16}, our optimized implementation achieves 1.1x$\sim$3.0x (2.15x on average) speedup on Tesla P100 with 17\% to 47.6\% (34.7\% on average) memory bandwidth utilization improvement compared to the GEMM function in cuBLAS.

\begin{figure}[t!]
    \centering
    \begin{subfigure}[t]{0.49\columnwidth}
        \includegraphics[width=\textwidth]{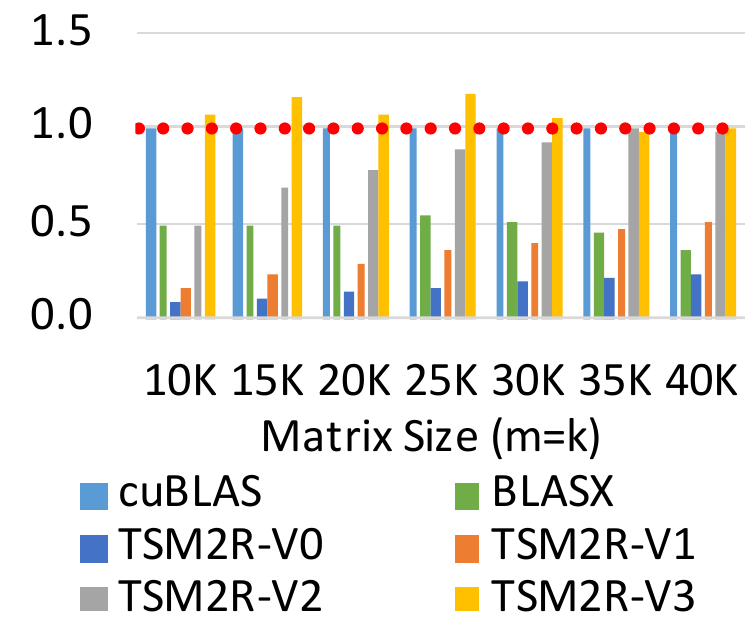}
        \caption{single (2)}
    \end{subfigure}
    \begin{subfigure}[t]{0.49\columnwidth}
        \includegraphics[width=\columnwidth]{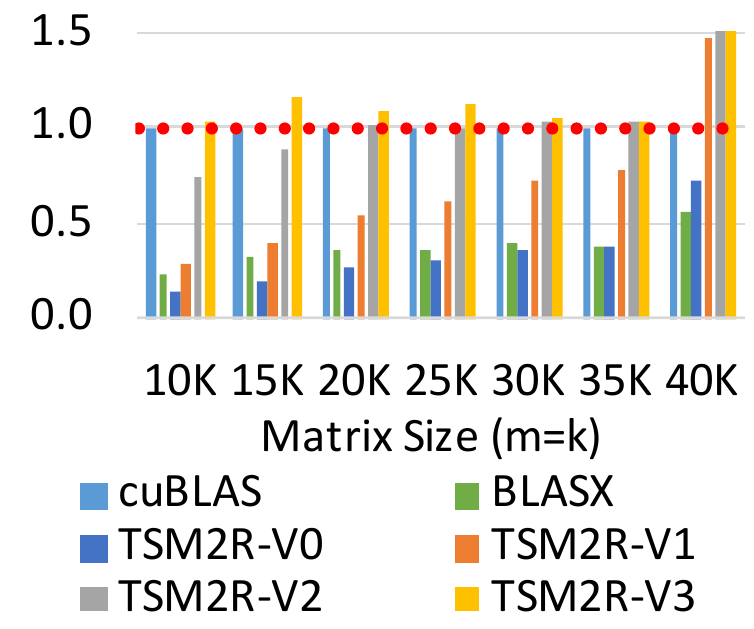}
        \caption{double (2)}
    \end{subfigure}
    \begin{subfigure}[t]{0.49\columnwidth}
        \includegraphics[width=\columnwidth]{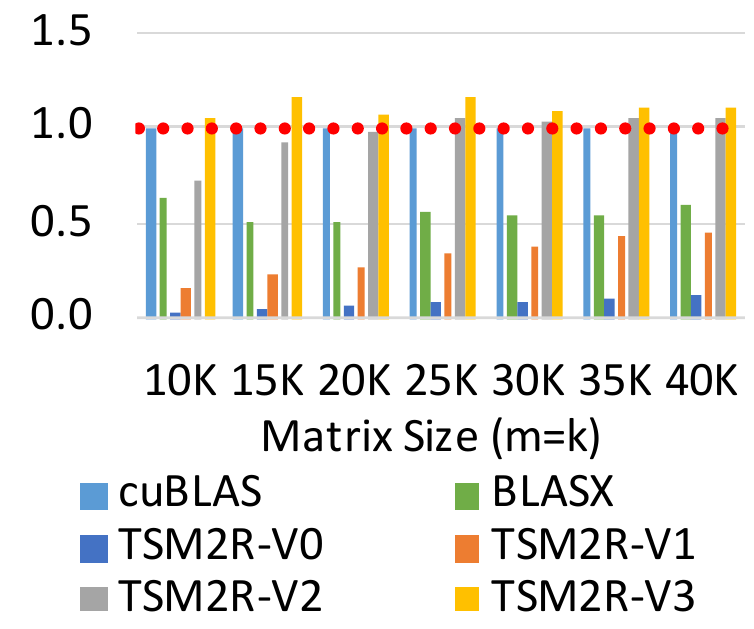}
        \caption{single (4)}
    \end{subfigure}
    \begin{subfigure}[t]{0.49\columnwidth}
        \includegraphics[width=\columnwidth]{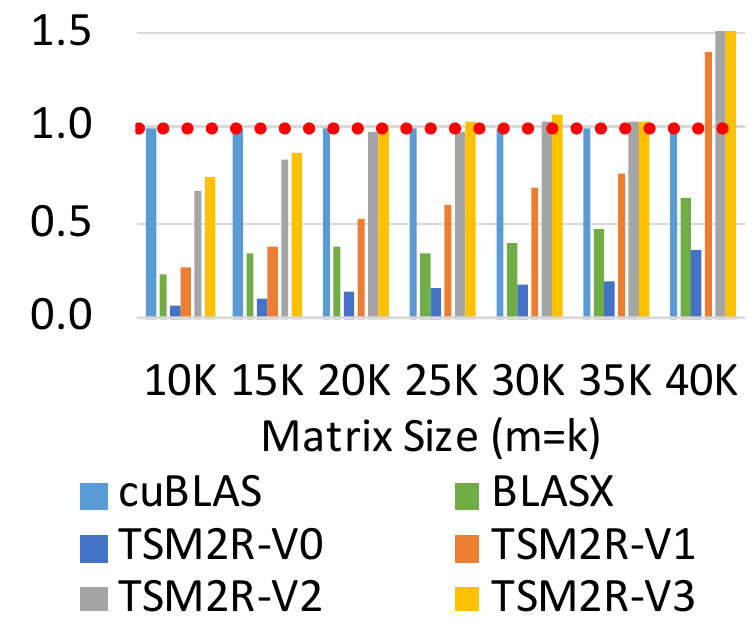}
        \caption{double (4)}
    \end{subfigure}
    \begin{subfigure}[t]{0.49\columnwidth}
        \includegraphics[width=\columnwidth]{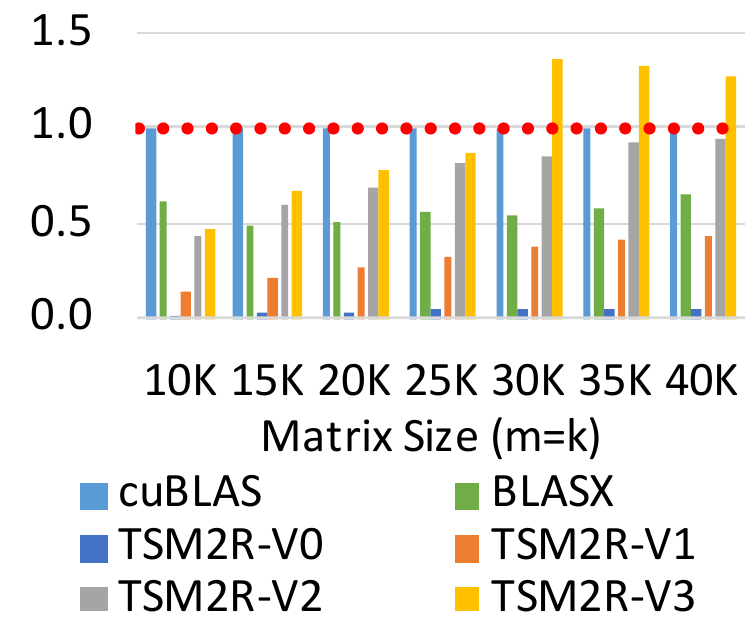}
        \caption{single (8)}
    \end{subfigure}
    \begin{subfigure}[t]{0.49\columnwidth}
        \includegraphics[width=\columnwidth]{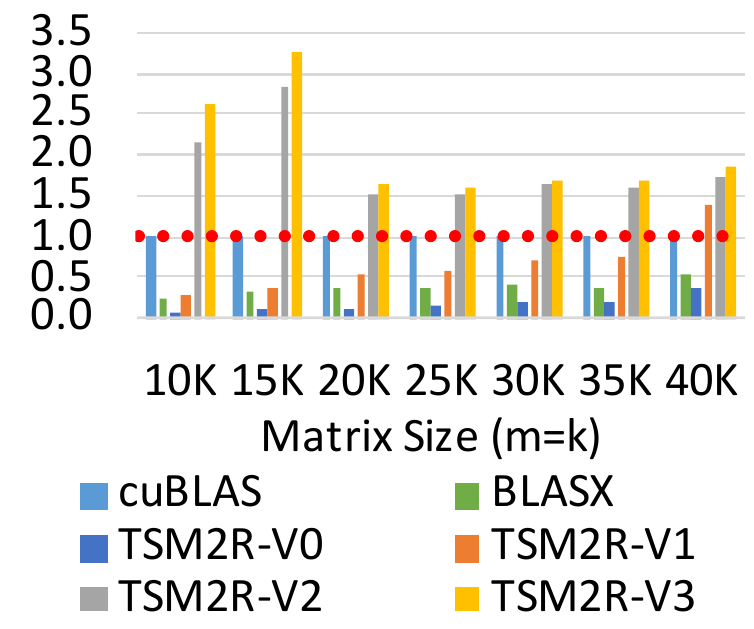}
        \caption{double (8)}
    \end{subfigure}
    \begin{subfigure}[t]{0.49\columnwidth}
        \includegraphics[width=\columnwidth]{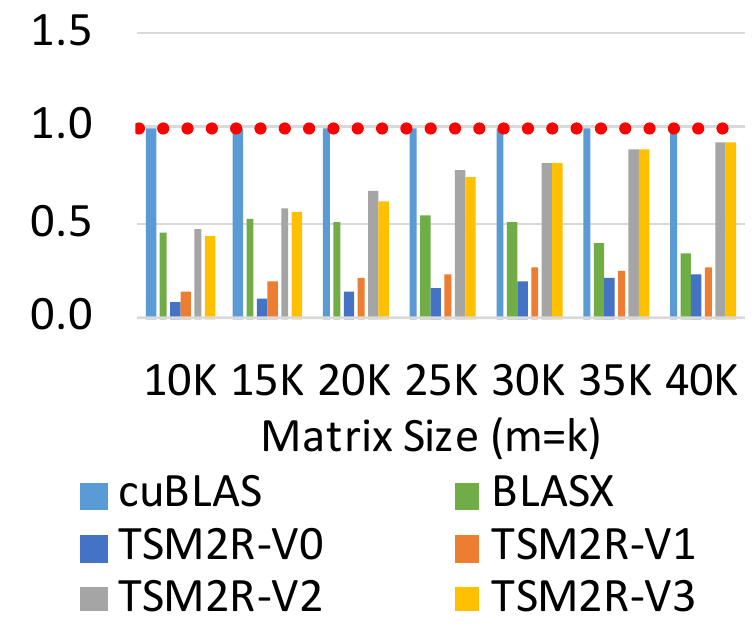}
        \caption{single (16)}
    \end{subfigure}
    \begin{subfigure}[t]{0.49\columnwidth}
        \includegraphics[width=\columnwidth]{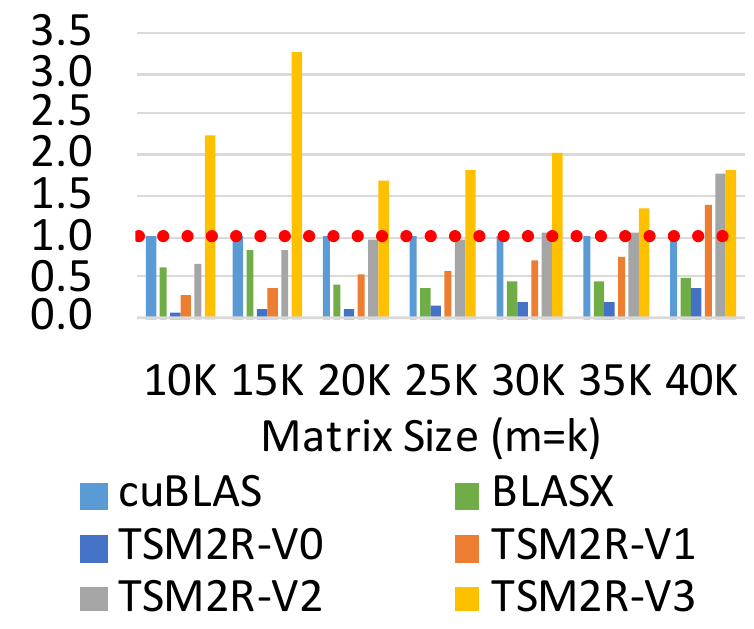}
        \caption{double (16)}
    \end{subfigure}
    \vspace*{-0.5em}
    \caption{Speedup comparison with single and double precision on V100 (n = 2, 4, 8, 16).}
    \label{v100-speedup}
%    \vspace*{-1.5em}
\end{figure}

\begin{figure}[t!]
    \centering
    \begin{subfigure}[t]{0.4\columnwidth}
        \includegraphics[width=\textwidth]{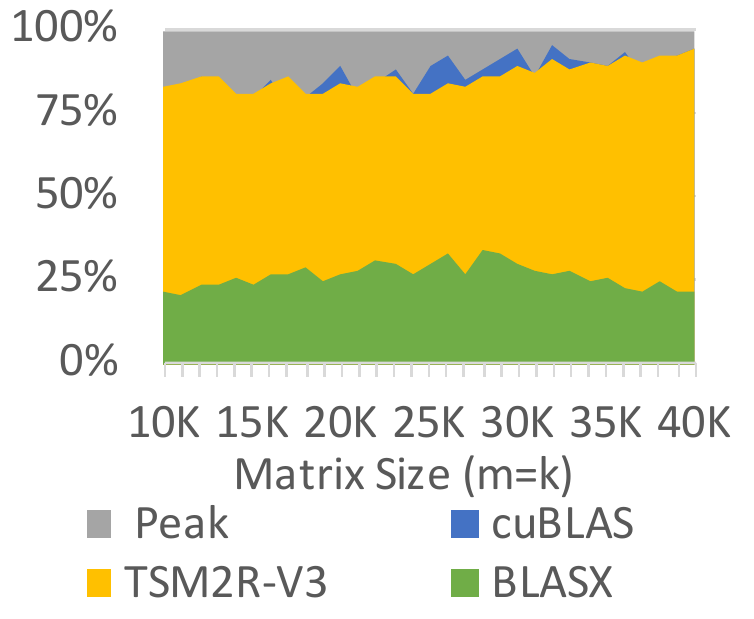}
        \caption{single (2)}
    \end{subfigure}
    \begin{subfigure}[t]{0.4\columnwidth}
        \includegraphics[width=\columnwidth]{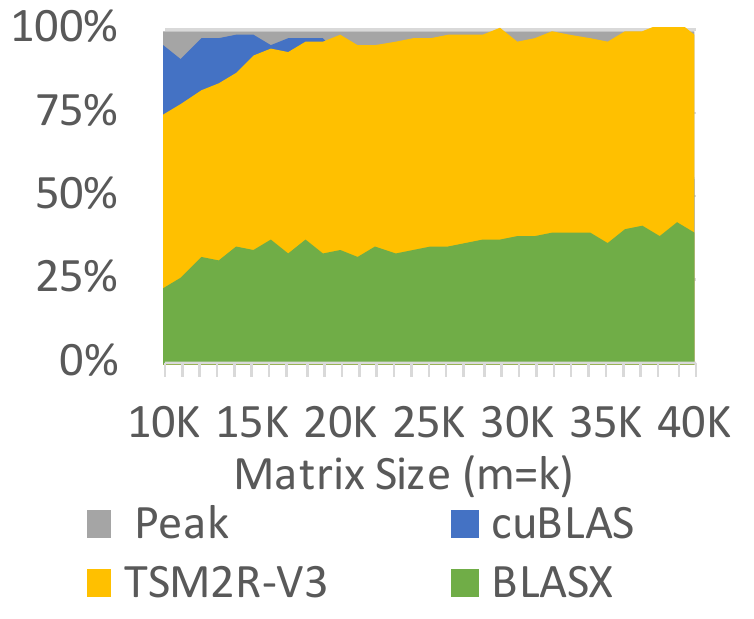}
        \caption{double (2)}
    \end{subfigure}
    \begin{subfigure}[t]{0.4\columnwidth}
        \includegraphics[width=\columnwidth]{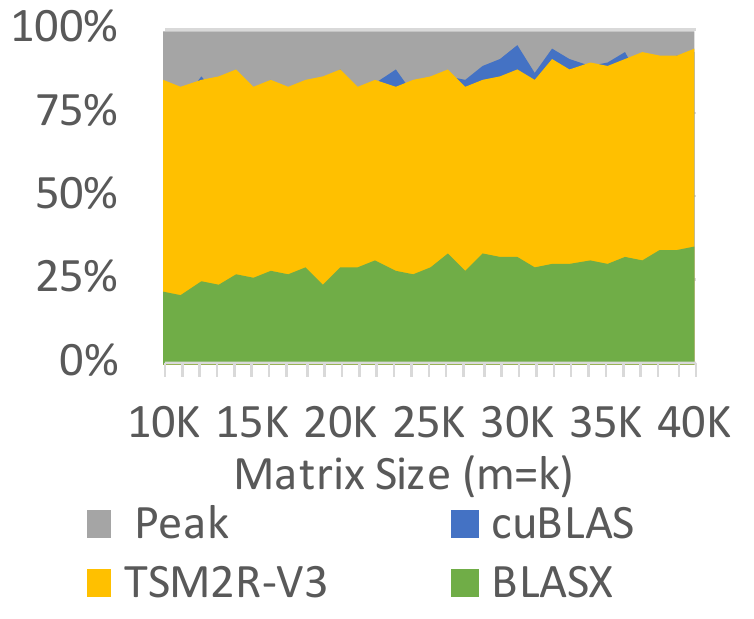}
        \caption{single (4)}
    \end{subfigure}
    \begin{subfigure}[t]{0.4\columnwidth}
        \includegraphics[width=\columnwidth]{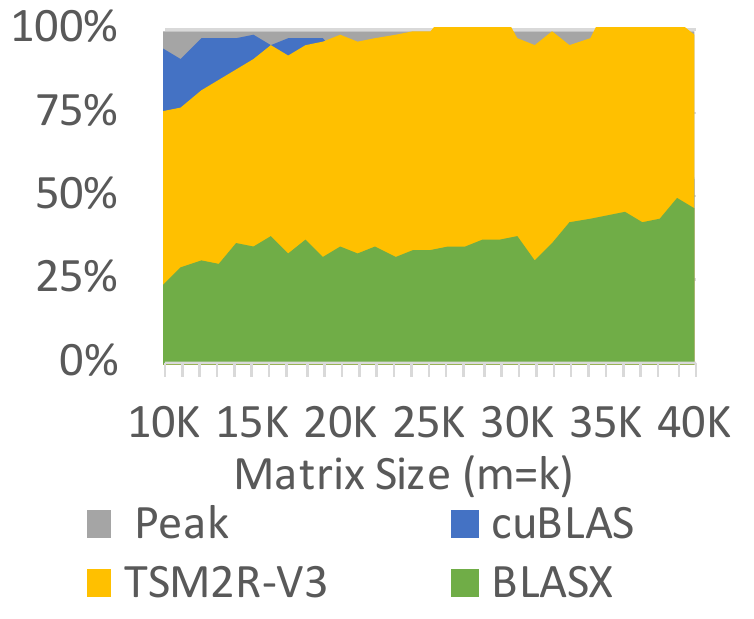}
        \caption{double (4)}
    \end{subfigure}
    \begin{subfigure}[t]{0.4\columnwidth}
        \includegraphics[width=\columnwidth]{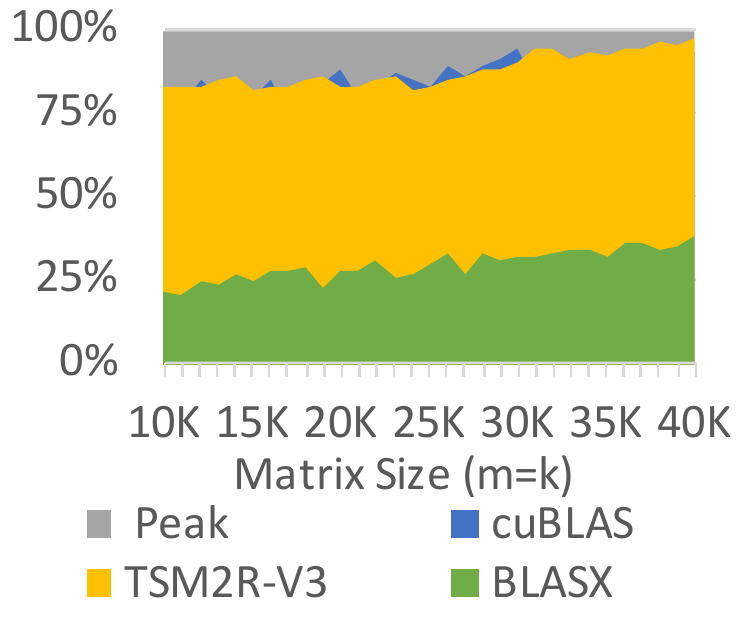}
        \caption{single (8)}
    \end{subfigure}
    \begin{subfigure}[t]{0.4\columnwidth}
        \includegraphics[width=\columnwidth]{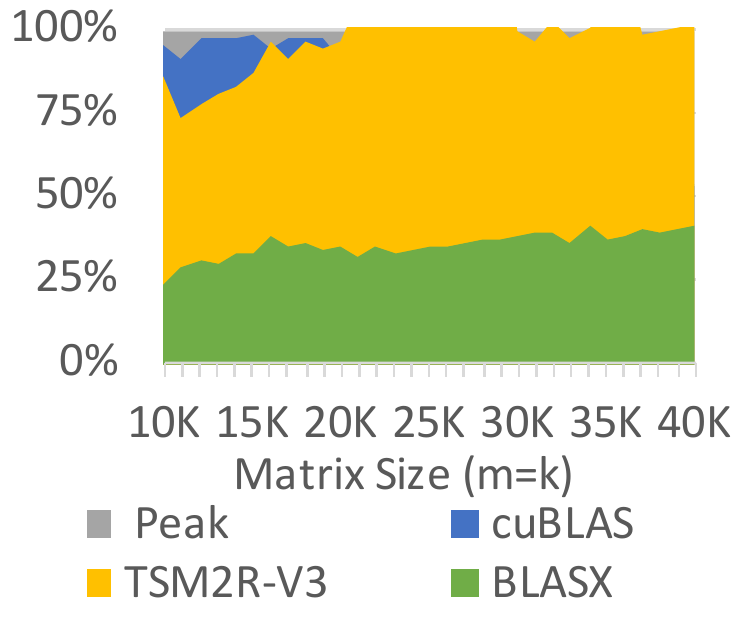}
        \caption{double (8)}
    \end{subfigure}
    \begin{subfigure}[t]{0.4\columnwidth}
        \includegraphics[width=\columnwidth]{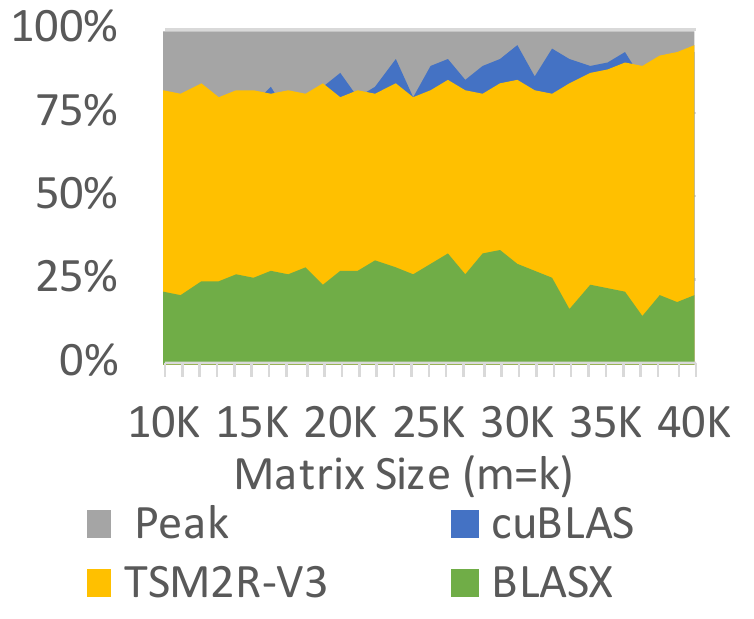}
        \caption{single (16)}
    \end{subfigure}
    \begin{subfigure}[t]{0.4\columnwidth}
        \includegraphics[width=\columnwidth]{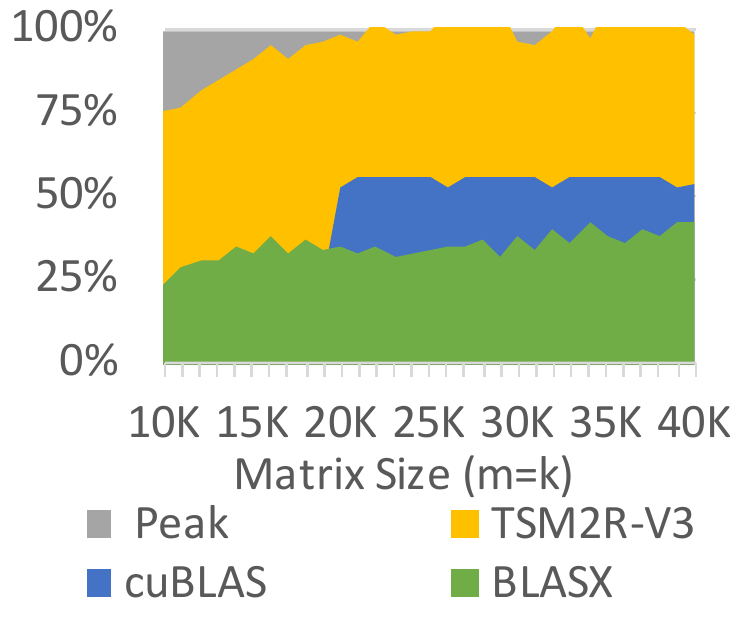}
        \caption{double (16)}
    \end{subfigure}
    \vspace*{-0.5em}
    \caption{Memory bandwidth utilization with single and double precision on V100 (n = 2, 4, 8, 16).}
    \label{v100-mem}
%    \vspace*{-2em}
\end{figure}

We also test \texttt{TSM2R} on the NVIDIA Tesla V100 GPU with Volta micro-architecture. Due to larger memory space on V100, we further increase the size of regular-shaped matrix to 40960 $\times$ 40960. \black{As $t_{2(v100)}^{threshold} \approx 70$, the computation is memory-bound for the given values. To ensure maximum performance and account for the Volta's architectural improvements, we optimize the parameters via brute-force. The optimized parameters are $t_1 = 128$, $t_2 = n$, and $t_3 = 32$ for single precision. For double precision, the optimized parameters are $t_1 = 128$, $t_2 = n$, and $t_3 = 16$ if $m < 10240$, or $t_3 = 12$ otherwise.} As shown in Figure \ref{v100-speedup}, we exhibit \black{gradually improving performance from \texttt{TSM2R-V0} to \texttt{TSM2R-V2}, similar to K40c. For \texttt{TSM2R-V3}, our best version,} speedups of up to 1.35x \black{(0.91x on average)} are achieved on single precision, while speedups of up to 3.2x \black{(1.5x on average)} are achieved on double precision. Note that the speedup for $n$ = $16$ on single precision is slower than cuBLAS. This is due to cuBLAS's single-precision GEMM being optimized for $32\times32$ matrices; thus we no longer target this case.
Finally, we note that the kernels achieve higher memory bandwidth utilization on V100 than on other GPUs, as shown in Figure \ref{v100-mem}. This is partly attributed to the improvements of Volta over previous micro-architectures. More specifically, V100 with improved HBM2 memory allows more workloads to obtain up to 19\% more memory bandwidth utilization than Pascal GPUs, according to its whitepaper \cite{voltawhitepaper}. \black{We provide experimental metrics for our \texttt{TSM2R} kernels in Table \ref{kerneldetails}.} 

\begin{table*}[ht]
\sffamily
\scriptsize
\centering
\begin{tabular}{rr|rrr|rrr}
      \textbf{GPU}
    & \textbf{Precision}
    & \textbf{$t_1$}
    & \textbf{$t_2$}
    & \textbf{$t_3$}
    & \textbf{Threads/Block}
    & \textbf{Shared Memory (bytes)}
    & \textbf{\# Registers}
    \\ \hline
      K40c
    & Single
    & 128
    & $n$
    & 4
    & 128
    & $512 \times (n \times 4)$
    & 64
    \\
      K40c
    & Double
    & 128
    & $n$
    & 4
    & 128
    & $1024 \times (n \times 8)$
    & 128
    \\
      M40
    & Single
    & 256
    & 8
    & 4
    & 256
    & 8192
    & 40
    \\
      M40
    & Double
    & 256
    & 8
    & 4
    & 256
    & 16384
    & 70
    \\
      P100
    & Single
    & 128
    & 4
    & 4
    & 128
    & 2048
    & 32
    \\
      P100
    & Double
    & 128
    & 4
    & 4
    & 128
    & 4096
    & 56
    \\
     V100
    & Single
    & 128
    & $n$
    & 32
    & 128
    & $512 \times (n \times 4)$
    & 144
    \\
     V100
    & Double
    & 128
    & $n$
    & 16
    & 128
    & $1024 \times (n \times 8)$
    & 180
    \\
     V100
    & Double
    & 128
    & $n$
    & 12
    & 128
    & $1024 \times (n \times 8)$
    & 168
    \\
\end{tabular}
\caption{Details of \texttt{TSM2R} kernel. Note that the number of registers is experimental data collected from NVCC and depends on $n$ ($n = 16$).}
\label{kerneldetails}
\end{table*}

\black{Due to our performance modeling, we can predict \texttt{TSM2R}'s performance on the Nvidia Tesla A100, with Ampere architecture. Its peak double-precision floating point performance is 9.7 TFLOPS, 1.3$\times$ that of the V100, and its global memory bandwidth is 1555 GB/s, 1.73$\times$ that of the V100. 
%For the sake of our calculations, we let the input be double precision and of the following size: $m = k = 15360$, $n = 16$. 
%Due to ???, we use $t_1 = ???$, $t_2 = ???$, $t_3 = ???$. 
Since all the cases considered in this paper are memory-bounded on the A100 (as $t_{2(a100)}^{threshold} \approx 50$) and our implementation has already achieved more than 90\% efficiency in memory bandwidth, we expect \texttt{TSM2R} can achieve a speedup of about 1.7 $\times$ on the A100 over the V100. Since we do not currently have access to a Nvidia Tesla A100 GPU, our estimates are based on the available whitepaper and do not take into account any other architectural improvements \cite{amperewhitepaper}.}

\subsubsection{Tests on Non-squared Input}

\begin{figure}[t!]
    \centering
    \begin{subfigure}[t]{0.45\columnwidth}
        \includegraphics[width=\textwidth]{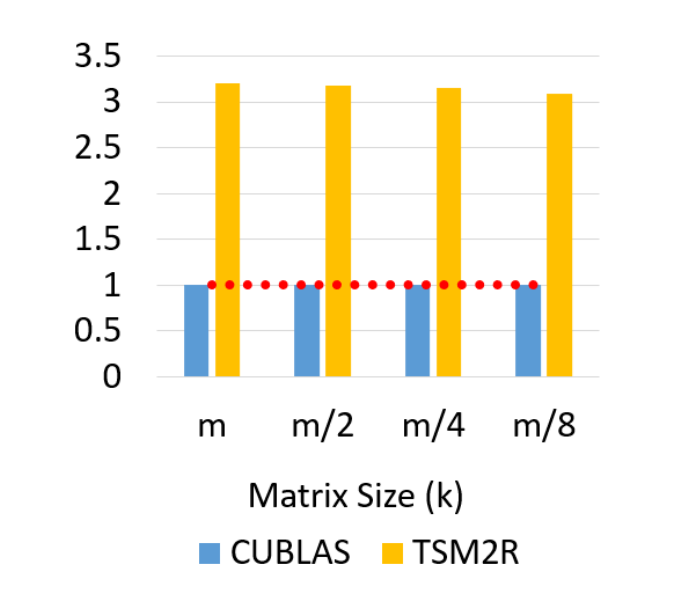}
        \caption{Speedup}
    \end{subfigure}
    \begin{subfigure}[t]{0.45\columnwidth}
        \includegraphics[width=\columnwidth]{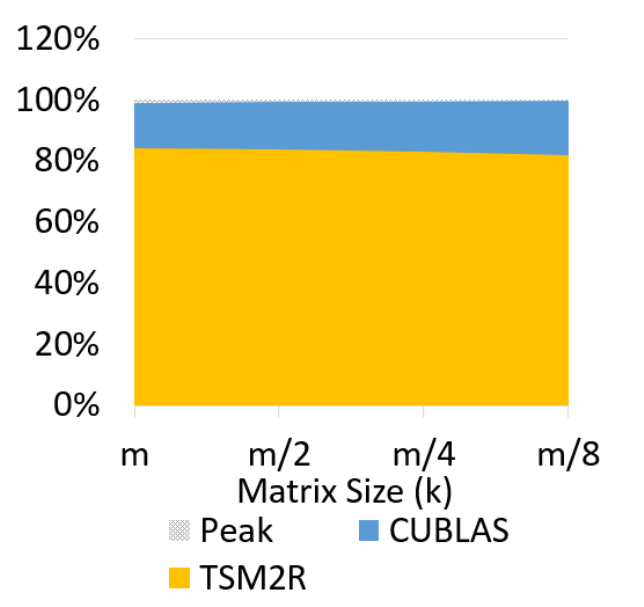}
        \caption{Mem. bw. util.}
    \end{subfigure}
    \vspace*{-0.5em}
    \caption{Performance comparison with double-precision rectangular input on V100 (m = 15360, n = 16).}
    \label{v100-varies}
%    \vspace*{-1.5em}
\end{figure}

We also evaluate \texttt{TSM2R} with rectangular input matrices ($m \times k$) on V100, where $k$ is smaller than $m$ by certain small integer factors. Evaluating this case reveals very little performance impact, as demonstrated in Figure \ref{v100-varies}. Although smaller than $m$, $k$ is still large enough to ensure the kernel to follow our performance model. The memory bandwidth utilization of the kernel remains similar to the case where $m = k$, and the performance of the kernel scales linearly with the matrix size.

\subsection{Evaluation of \texttt{TSM2L}}

\begin{figure}[t!]
    \centering
    \begin{subfigure}[t]{0.49\columnwidth}
        \includegraphics[width=\columnwidth]{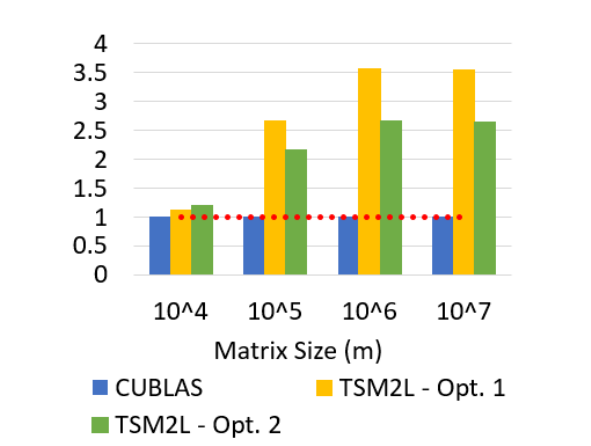}
        \caption{single (8)}
    \end{subfigure}
    \begin{subfigure}[t]{0.49\columnwidth}
        \includegraphics[width=\columnwidth]{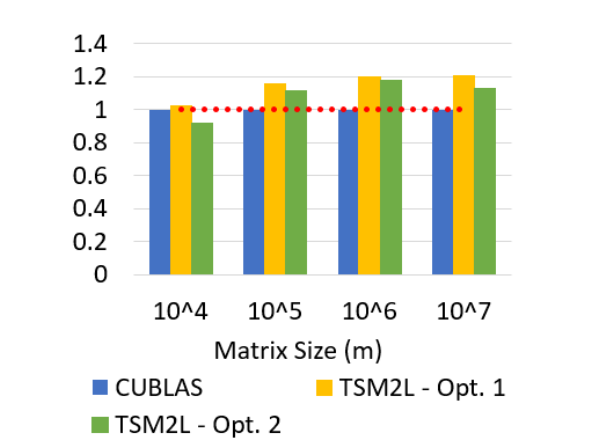}
        \caption{double (8)}
    \end{subfigure}
    \vspace*{-0.5em}
    \begin{subfigure}[t]{0.49\columnwidth}
        \includegraphics[width=\columnwidth]{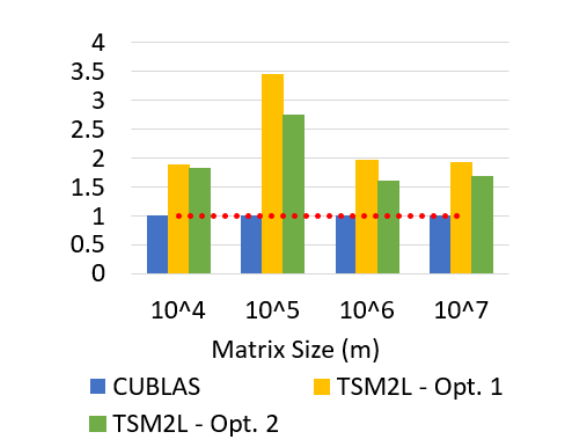}
        \caption{single (16)}
    \end{subfigure}
    \begin{subfigure}[t]{0.49\columnwidth}
        \includegraphics[width=\columnwidth]{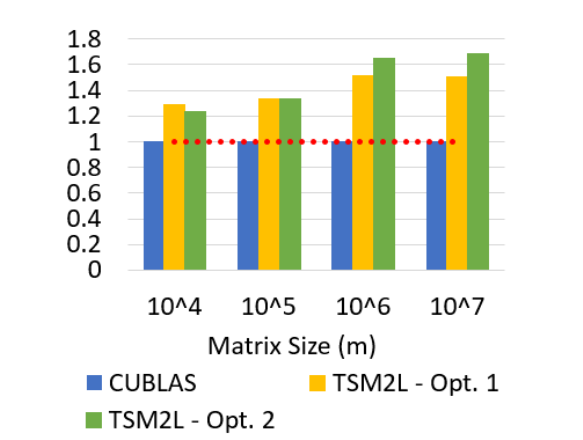}
        \caption{double (16)}
    \end{subfigure}
    \vspace*{-0.5em}
    \caption{Speedup comparison with single and double precision on V100 (k = n = 8, 16).}
    \label{ism2-perf}
%   \vspace*{-0.5em}
\end{figure}

\begin{figure}[t!]
     \centering
    \begin{subfigure}[t]{0.49\columnwidth}
        \includegraphics[width=\columnwidth]{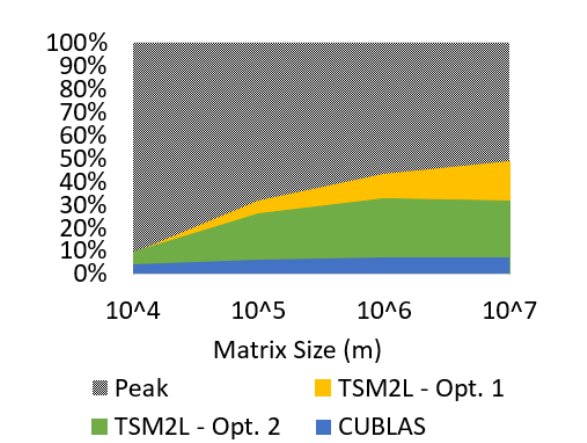}
        \caption{single (8)}
    \end{subfigure}
    \begin{subfigure}[t]{0.49\columnwidth}
        \includegraphics[width=\columnwidth]{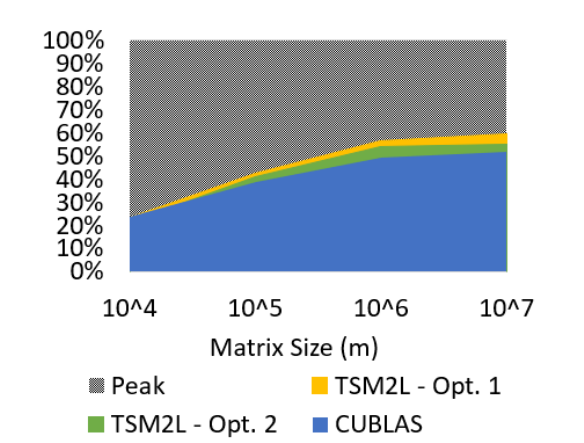}
        \caption{double (8)}
    \end{subfigure}
    \begin{subfigure}[t]{0.49\columnwidth}
        \includegraphics[width=\columnwidth]{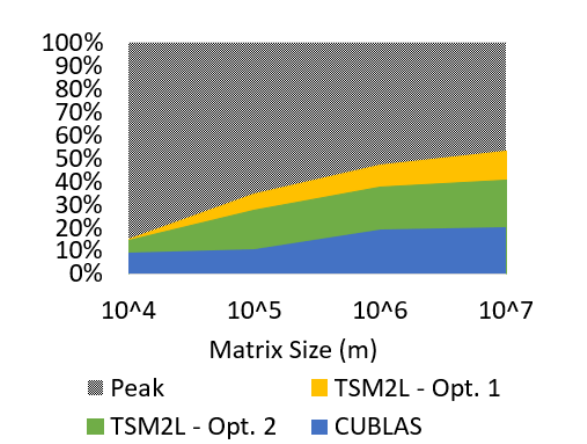}
        \caption{single (16)}
    \end{subfigure}
    \begin{subfigure}[t]{0.49\columnwidth}
        \includegraphics[width=\columnwidth]{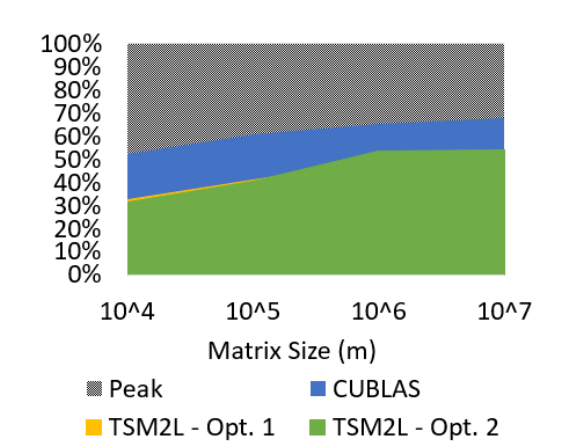}
        \caption{double (16)}
    \end{subfigure}
    \vspace*{-0.5em}
    \caption{Memory bandwidth utilization with single and double precision on V100 (k = n = 8, 16).}
    \label{ism2-mem}
%    \vspace*{-0.5em}
\end{figure}

\begin{table*}[!ht]
\sffamily
\scriptsize
\centering
\begin{tabular}{rr|rrr|rrr}
      \textbf{Optimization}
    & \textbf{Precision}
    & \textbf{$t_1$}
    & \textbf{$t_2$}
    & \textbf{$t_3$}
    & \textbf{Threads/Block}
    & \textbf{Shared Memory (bytes)}
    & \textbf{\# Registers}
    \\ \hline
      \texttt{TSM2L-Opt1}
    & Single
    & 128
    & $n$
    & 32
    & 128
    & $512 \times (n \times 4)$
    & 144
    \\
      \texttt{TSM2L-Opt1}
    & Double
    & 128
    & $n$
    & 16
    & 128
    & $1024 \times (n \times 8)$
    & 180
    \\
     \texttt{TSM2L-Opt1}
    & Double
    & 128
    & $n$
    & 12
    & 128
    & $1024 \times (n \times 8)$
    & 168
    \\
      \texttt{TSM2L-Opt2}
    & Single
    & 128
    & $n$
    & 32
    & 128
    & $512 \times (n \times 4)$
    & 251
    \\
     \texttt{TSM2L-Opt2}
    & Double
    & 128
    & $n$
    & 16
    & 128
    & $1024 \times (n \times 8)$
    & 254
    \\
     \texttt{TSM2L-Opt2}
    & Double
    & 128
    & $n$
    & 12
    & 128
    & $1024 \times (n \times 8)$
    & 252
    \\
    % \bottomrule
\end{tabular}
\caption{Details of \texttt{TSM2L} kernel on V100. Note that the number of registers is experimental data collected from NVCC and depends on $n$ ($n = 16$).}
\label{kerneldetailsl}
\end{table*}

We next evaluate the performance of \texttt{TSM2L} and compare it with cuBLAS on V100. For \texttt{TSM2L}, we must choose the variable $tcf$ for each matrix input combination. We obtain these values through experiments that vary $tcf$. As a result, for $m = 10^4, 10^5, 10^6,$ $10^7$, we select $tcf$ values as 1, 1, 2, and 8 for single precision and values 1, 1, 1, and 4 for double precision. Considering two proposed optimizations for \texttt{TSM2L}, we have two versions of \texttt{TSM2L}\black{: \texttt{TSM2L-Opt1}, based on Algorithm \ref{alg-opt1}, and \texttt{TSM2L-Opt2}, based on Algorithm \ref{alg-opt2}.}

As shown in Figure \ref{ism2-perf}, \texttt{TSM2L} can obtain speedups over cuBLAS ranging from 1.1x$\sim$3.5x \black{(2.5x on average)} in single precision and speedups from 1.0x$\sim$1.7x \black{(1.3x on average)} in double precision. \texttt{TSM2L-Opt1} generally performs better on single precision input than \texttt{TSM2L-Opt2}, while \texttt{TSM2L-Opt2} performs better than \texttt{TSM2L-Opt1} in several double precision cases. In addition, as shown in Figure \ref{ism2-mem}, \texttt{TSM2L} achieves memory bandwidth utilization of up to 55\% peak global memory bandwidth \black{(40\% on average)}. In single precision, \texttt{TSM2L} utilizes significantly more memory bandwidth than cuBLAS, up to 40\% more \black{(25\% on average)}. However, in double precision, \texttt{TSM2L} uses only slightly more memory bandwidth in the case that $k$ = $n$ = $8$, and in the case that $k$ = $n$ = $16$, cuBLAS uses more memory bandwidth. However, since \texttt{TSM2L} still outperforms cuBLAS, this can be explained by inefficient memory use patterns in the GEMM kernel. \black{We provide experimental metrics for our \texttt{TSM2L} kernels in Table \ref{kerneldetailsl}.} 
\section{Related Works}
\label{related}
\black{
A preliminary version of this work was published in \cite{chen2019tsm2}. It introduces our \texttt{TSM2R} algorithm and evaluates it on Kepler, Maxwell, and Pascal GPUs respectively. In this paper, we expand the evaluation by adding experiments on the Volta GPU V100. Moreover, we also broaden the applicability of this work through our new \texttt{TSM2L} algorithm to handle a new input case.
}

\black{
Ernst et al.'s work also focuses on optimizing tall-and-skinny GEMM \cite{ernst2020perf}. It proposes two algorithms for tall-and-skinny input: \texttt{TSMTTSM}, where a tall-and-skinny matrix is multiplied by transposed tall-and-skinny matrix, and \texttt{TSMM}, where a tall-and-skinny matrix is multiplied by a small square matrix. It evaluates these algorithms on the Volta GPU V100 with double-precision real and complex floating-point numbers. 
Although \texttt{TSMTTSM}'s input case is not considered by our work, \texttt{TSMM}'s input case is the same as \texttt{TSM2L}'s. However, \texttt{TSMM} and \texttt{TSM2L} approach this input case differently. Specifically, \texttt{TSMM} takes matrices in row-major format as input, so its optimizations focus on avoiding partially-written cache lines while storing columns of matrix C. \texttt{TSMM} launches multiple threads per row of matrix C, with each thread storing several columns of matrix C. Unlike \texttt{TSMM}, the design of our \texttt{TSM2L} takes into account the latency of launching many threads that perform little work, and its optimizations' focus on managing both warp and memory bandwidth latency. Moreover, note that our \texttt{TSM2L} achieves superior performance. \texttt{TSMM} only achieves speedups over cuBLAS where the small dimension, $k = n$, is less than $8$, whereas our \texttt{TSM2L} achieves speedups with the small dimension up to $16$.
}
\section{Conclusion}
\label{conclusion}
In this work, we first analyze the performance bottleneck of current GEMM in the latest cuBLAS library. We identify that current implementations lack of full utilization of computing power or memory bandwidth when the input shape is tall-and-skinny. Then, we discovered the potential challenges of optimizing  tall-and-skinny GEMM since its workload can vary between compute bound, memory bound, and latency bound. Next, we propose two high-performance GEMM algorithms---\texttt{TSM2R} and \texttt{TSM2L}---on GPUs for tall-and-tinny input with several optimization techniques focusing on GPU resource utilization. Finally, experiment results show that our optimized implementations can achieve speedups tall-and-skinny matrix-matrix multiplication with diverse input sizes on modern GPUs.
\section*{Acknowledgements}
This research is supported by the National Science Foundation under Grants OAC-2034169 and OAC-2003624. We would like to thank the University of Alabama for providing the startup support in this work. The authors acknowledge the Texas Advanced Computing Center (TACC) at The University of Texas at Austin for providing HPC resources that have contributed to the research results reported within this paper.

\bibliographystyle{elsarticle-num}
\bibliography{ref}

\newpage

\parpic{\includegraphics[width=1in,clip,keepaspectratio]{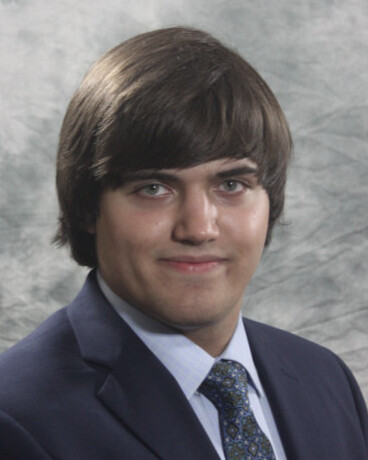}}
\noindent {\bf Cody Rivera} is an undergraduate student studying Computer Science and Mathematics at the University of Alabama from Fall 2018. He is also in the Randall Research Scholars Program, an honors interdisciplinary undergraduate research program. His research interests include computer science theory, algorithms, and high-performance computing.

\parpic{\includegraphics[width=1in,clip,keepaspectratio]{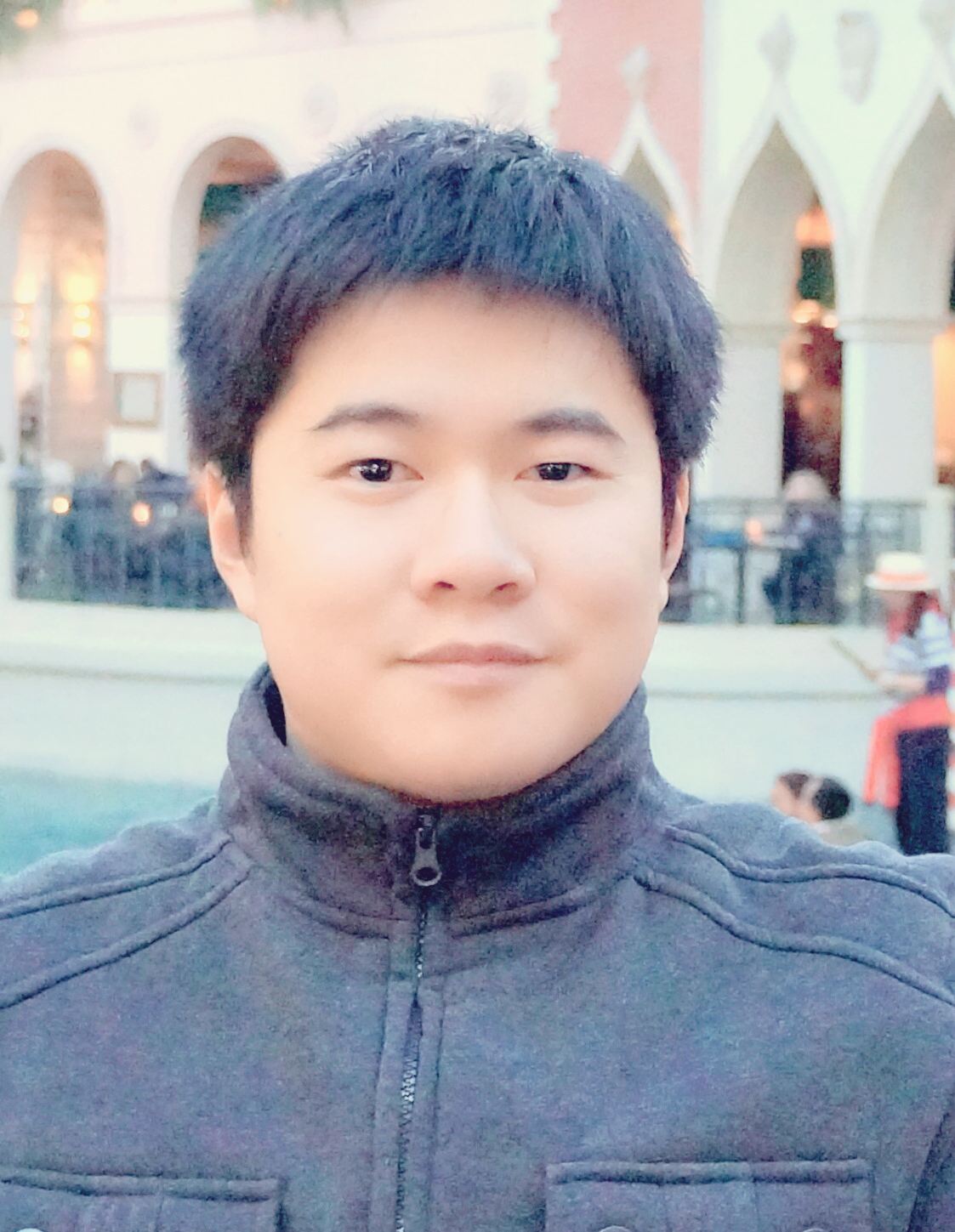}}
\noindent {\bf Jieyang Chen} is a postdoctoral researcher in the Computer Science and Mathematics Division at Oak Ridge National Laboratory (ORNL). He received his master's and Ph.D. degrees in Computer Science from University of California, Riverside in 2014 and 2019. He received a bachelor's degree in Computer Science and Engineering from Beijing University of Technology in 2012. Before joining ORNL, he interned at Pacific Northwest National Laboratory and Los Alamos National Laboratory. His research interests include high-performance computing, parallel and distributed systems, and big data analytics. He has published over 20 peer-reviewed high-quality papers in prestigious HPC and Big Data conferences and journals, such as ICS, HPDC, PPoPP, SC, BigData, IPDPS, TPDS.

\parpic{\includegraphics[width=1in,clip,keepaspectratio]{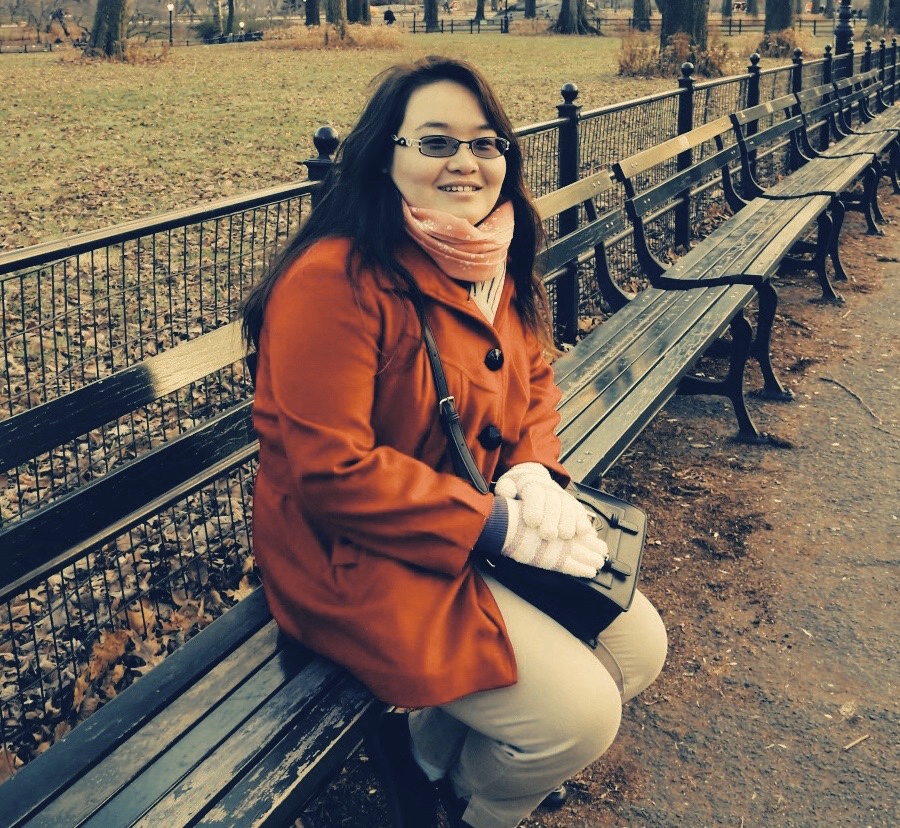}}
\noindent {\bf Nan Xiong} graduated with a master’s degree in Computer Science from University of California, Riverside in 2018. She also received a master’s degree in Civil Engineering from University of Southern California in 2014 and a bachelor’s degree in Civil Engineering from Tianjin University in 2012. She is interested in HPC, heterogeneous computing with GPU accelerators, and high-performance big data analytics. 

\parpic{\includegraphics[width=1in,clip,keepaspectratio]{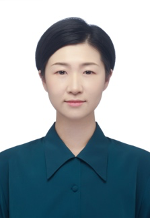}}
\noindent {\bf Jing Zhang} obtained her master's degree in Business Administration and Management from University of Colorado in 2015. She has been selected for membership in Beta Gamma Sigma, the international honor society for collegiate schools of business. She was International Development Advisor in FDI Strategies. She is currently an independent researcher in undergraduate college of Wuhan University, focusing on business data analysis and information system design. 

\parpic{\includegraphics[width=1in,clip,keepaspectratio]{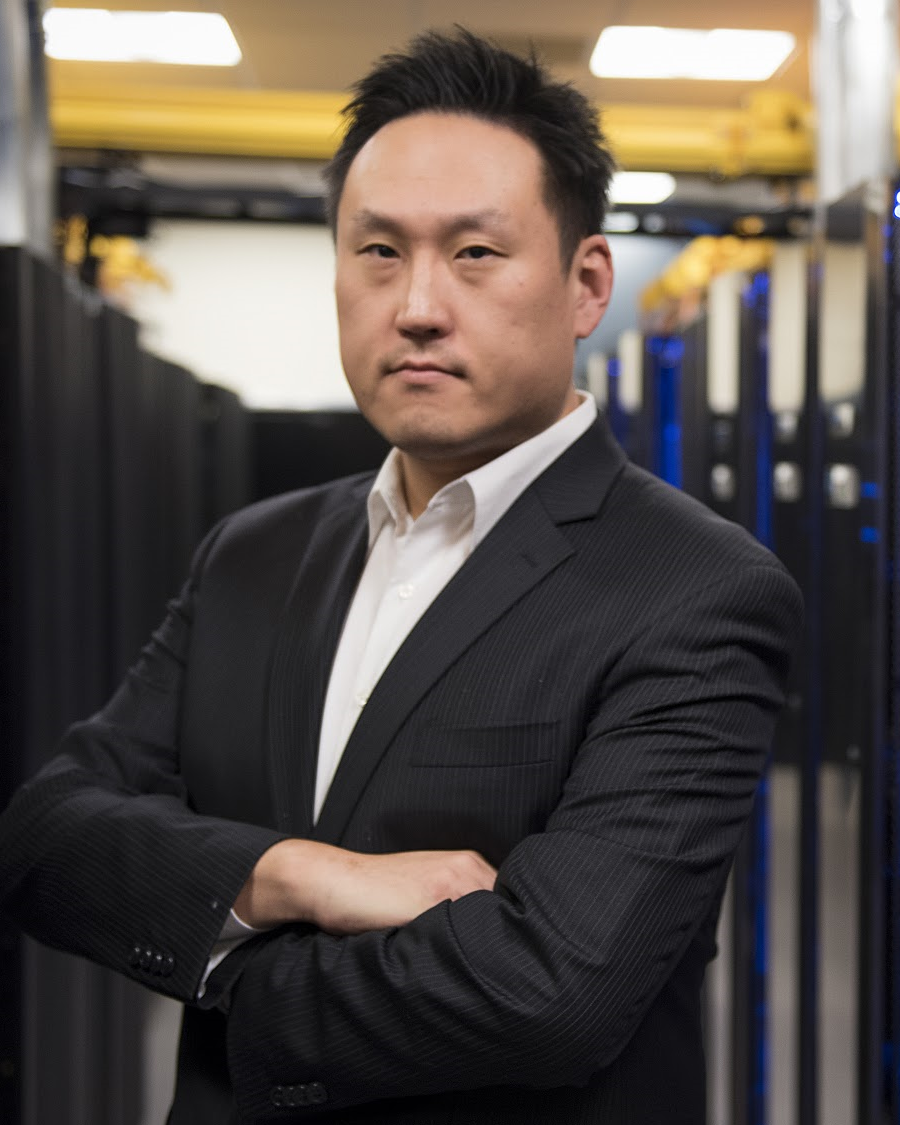}}
\noindent {\bf Shuaiwen Leon Song} is currently a senior lecturer (tenured associate professor) at the school of computer science of University of Sydney and the director of Future System Architecture Lab. He is affiliated with USYD nanoscience hub and Sydney Quantum Academy. He is also an affiliated professor with University of Washington. Prior to his appointment at University of Sydney, he worked for U.S. Department of Energy as a senior research scientist and technical lead. His research interests include holistic system design, system architecture and high performance computing. His most recent works target future accelerator-driven system design for AI and planet-scale virtual reality. He is a Lawrence scholar, Paul E. Torgersen scholar, a recipient of IEEE TCHPC early career award and DOE pathway to excellence research award. He widely published in the major HPC and computer architecture conferences, including ISCA, HPCA, MICRO, ASPLOS and SC. His past work received a 2017 HiPEAC paper award, two SC best paper runner-ups, and 2018 IISWC best paper finalist. During his tenure at PNNL, he led two LDRD projects on AI driven future HPC system design and large-scale data analytics acceleration. 

\parpic{\includegraphics[width=1in,clip,keepaspectratio]{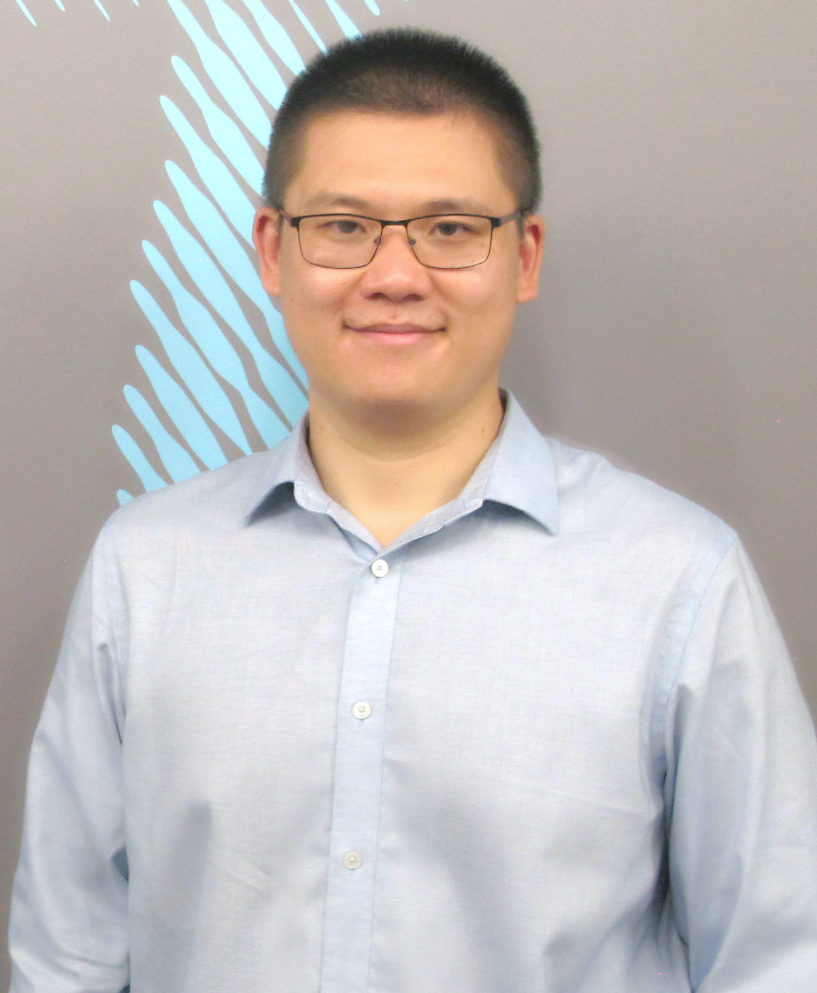}}
\noindent {\bf Dingwen Tao} is an assistant professor in the School of Electrical Engineering and Computer Science at Washington State University. Prior to that, he worked as an assistant professor at the University of Alabama, and interned at Brookhaven National Laboratory, Argonne National Laboratory, and Pacific Northwest National Laboratory. He received his Ph.D. in Computer Science from University of California, Riverside in 2018 and B.S. in Mathematics from University of Science and Technology of China in 2013. He currently works at the intersection of HPC and big data analytics, focusing on scientific data management, HPC storage and I/O systems, fault tolerance at extreme scale, and distributed machine learning. He has published in major HPC and big data analytics conferences and journals, such as SC, PPoPP, HPDC, ICS, PACT, IPDPS, Cluster, ICPP, BigData, and TPDS. He is the receipt of the IEEE CS TCHPC Early Career Researchers Award for Excellence in High Performance Computing, NSF CISE Research Initiation Initiative (CRII) Award in 2020, and UCR Dissertation Year Program Award in 2017.
\end{document}